\documentclass[reqno]{amsart}
\usepackage{amssymb}
\usepackage{upref}

\newcommand{\bbN}{{\mathbb{N}}}
\newcommand{\bbR}{{\mathbb{R}}}

\newcommand{\bbZ}{{\mathbb{Z}}}
\newcommand{\bbC}{{\mathbb{C}}}

\newcommand{\calB}{{\mathcal B}}

\newcommand{\calC}{{\mathcal C}}

\newcommand{\calD}{{\mathcal D}}

\newcommand{\calM}{{\mathcal M}}
\newcommand{\calE}{{\mathcal E}}
\newcommand{\calL}{{\mathcal L}}
\newcommand{\calF}{{\mathcal F}}
\newcommand{\calK}{{\mathcal K}}

\newcommand{\calU}{{\mathcal U}}
\newcommand{\bbc}{{\mathbb{C}}}
\newcommand{\N}{n}

\newcommand{\dott}{\,\cdot\,}
\newcommand{\hatt}{\widehat}  

\newcommand{\Div}{\operatorname{Div}}

\newcommand{\no}{\nonumber}
\newcommand{\lb}{\label}
\newcommand{\f}{\frac}
\newcommand{\ul}{\underline}
\newcommand{\ol}{\overline}
\newcommand{\ti}{\widetilde} 
\newcommand{\Oh}{O}

\newcommand{\lam}{\lambda}
\newcommand{\humu}{{ \hat{\underline{\mu} }}}
\newcommand{\hunu}{{\underline{\hat{\nu}}}}
\newcommand{\huunu}{\underline{\underline{\hat{\nu}}}}

\newcommand{\hmu}{{\hat{\mu} }}
\newcommand{\hnu}{{\hat{\nu}}}
\newcommand{\uz}{{\underline{z}}}

\newcommand{\hulam}{{ \hat{\underline{\lam} }}}
\newcommand{\hlam}{{ \hat{\lam} }}

\newcommand{\uxi}{{\underline{\Xi}}}
\newcommand{\ual}{{\underline{\alpha}}}
\newcommand{\ua}{{\underline{A}}}
\newcommand{\Pinfp}{{P_{\infty_+}}}
\newcommand{\Pinfm}{{P_{\infty_-}}}
\newcommand{\Pzerop}{{P_{0,+}}}
\newcommand{\Pzerom}{{P_{0,-}}}
\newcommand{\Pinfpm}{{P_{\infty_\pm}}}
\newcommand{\Pzeropm}{{P_{0,\pm}}}
\newcommand{\Pinfmp}{{P_{\infty_\mp}}}
\newcommand{\Pzeromp}{{P_{0,\mp}}}
\newcommand{\uu}{\bar u}
\newcommand{\vv}{\bar v}

\renewcommand{\Im}{\text{\rm Im}}

\renewcommand{\log}{\text{\rm log}}

\allowdisplaybreaks
\numberwithin{equation}{section}

\newtheorem{theorem}{Theorem}[section]
\newtheorem{lemma}[theorem]{Lemma}

\newtheorem{hypothesis}[theorem]{Hypothesis}
\theoremstyle{definition}

\newtheorem{example}[theorem]{Example}

\theoremstyle{remark}
\newtheorem{remark}[theorem]{Remark}

\newcommand{\abs}[1]{\lvert#1\rvert}

\begin{document}
\title[Thirring system]{The classical massive Thirring system revisited}
\author[Enolskii]{V.\ Z.\  Enolskii}
\address{Institute of Magnetism NASU, Vernadsky str. 36, Kiev 252142,
Ukraine}
\email{vze@imag.kiev.ua}
\thanks{The research of the first author was supported in part by 
the CRDF grant UM1-325, INTAS-96-770,  
Ex-Quota Royal Society grant, and the Norwegian Research Council.}
\author[Gesztesy]{F.\ Gesztesy}
\address{Department of Mathematics,
University of Missouri,
Columbia, MO 65211, USA}
\email{fritz@math.missouri.edu}
\urladdr{http://www.math.missouri.edu/people/fgesztesy.html}
\thanks{The research of the second author was supported in part by 
the CRDF grant UM1-325, the University of Missouri Research Board 
grant RB-97-086, and the Norwegian Research Council.}
\author[Holden]{H.\ Holden}
\address{Department of Mathematical Sciences,
Norwegian University of
Science and Technology, N--7491 Trondheim, Norway}
\email{holden@math.ntnu.no}
\urladdr{http://www.math.ntnu.no/\~{}holden/}
\thanks{The research of the third author was supported in part by 
the Norwegian Research Council.}
\dedicatory{Dedicated with great pleasure to Sergio Albeverio on the
occasion of his 60th birthday}

\begin{abstract}
We provide a complete treatment of algebro-geometric solutions of the
classical massive Thirring system.  In particular, we study
Dubrovin-type equations for auxiliary divisors, consider the
corresponding algebro-geometric initial value problem, and derive the
theta function representations of algebro-geometric solutions of the
Thirring system.
\end{abstract}

\maketitle

\section{Introduction}\lb{Ts1}

Ever since its publication in 1958, the Thirring model 
\cite{Thirring:1958} kept its fascination as is witnessed by the 
incredible amount of attention paid to it since then and by the
interest  it continues to generate (see, e.g., \cite{IlievaThirring:1998}
for a  recent review).  In the present paper we are not concerned with its 
importance as a solvable quantum field theory model but rather 
restrict our attention to its complete integrability aspects from a 
classical point of view.  Thirring's classical $(1+1)$-dimensional 
model equations in appropriate light cone coordinates, and after 
appropriate rescaling of the mass and coupling constant parameters, 
etc., can be cast in the form
\begin{align}
-iu_{x}+2v+2\abs{v}^2u&=0, \no \\
-iv_{t}+2u+2\abs{u}^2v&=0. \lb{T1.1}
\end{align}
Formal integrability of \eqref{T1.1} was originally established by 
Mikhailov \cite{Mikhailov:1976} in 1976 by establishing a 
corresponding commutator representation (cf.\ \eqref{T2.8} and 
\eqref{T2.9}).  In fact, one can replace \eqref{T1.1} by a more 
general system, without identifying $u^{*}$ and  $v^{*}$ with the 
complex conjugates  $\uu$ and  $\vv$ of $u$ and  $v$, respectively, 
\begin{align}
-iu_{x}+2v+2v v^*u&=0, \no \\
iu^{*}_{x}+2v^{*}+2v v^*u^{*}&=0,\lb{T1.2} \\
-iv_{t}+2u+2u u^{*}v&=0, \no \\
iv^{*}_{t}+2u^{*}+2u u^{*}v^{*}&=0, \no
\end{align}
without losing formal integrability, and we will actually investigate 
\eqref{T1.2} rather than \eqref{T1.1}. Both \eqref{T1.1} and 
\eqref{T1.2} have been studied by numerous authors, who derived the 
inverse scattering approach \cite{KaupLakoba:1996}, 
\cite{KawataMorishimaInoue:1979}, \cite{KuznetsovMikhailov:1977}, 
\cite{Villarroel:1991}, considered soliton solutions
\cite{BarashenkovGetmanov:1987},  
\cite{BarashenkovGetmanovKovtun:1993}, \cite{BarashenkovGetmanov:1993}, 
 \cite{Date:1979}, \cite{Date:1982}, 
\cite{DavidHarnadShnider:1984},  \cite{Talalov:1987}, 
\cite{Vaklev:1996}, investigated B\"acklund transformations and close 
connections with other integrable equations (especially, the 
sine-Gordon equation) \cite{Martinezalonso:1984}, \cite{KaupNewell:1977}, 
\cite{Lee:1993}, \cite{Lee:1994},  
\cite{NijhoffCapelQuispelLinden:1983}, \cite{Prikarpatskii:1981}, 
\cite{PrikarpatskiiGolod:1979}, \cite{TsuchidaWadati:1996}, 
\cite{WadatiSogo:1983}, and considered monodromy deformations 
\cite{ChowdhuryNaskar:1988}.

In the present paper we focus on algebro-geometric solutions of the 
classical massive Thirring system \eqref{T1.2}.  The first attempt to 
derive algebro-geometric solutions of \eqref{T1.1} is due to Date 
\cite{Date:1978} in 1978 and almost simultaneously to Prikarpatskii 
and Holod \cite{PrikarpatskiiGolod:1979} (see also 
\cite{HolodPrikarpatsky:1978}). Both papers are remarkably similar 
in strategy, in fact, they are nearly identical. In particular, both 
discuss theta function representations for symmetric functions of 
appropriate symmetric functions associated with auxiliary divisors, 
but neither derives explicit theta function representations of $u$ 
and $v$.  
The first theta function representations of $u$, $v$, $u^{*}$, 
$v^{*}$ for the general massive Thirring system \eqref{T1.2} were 
derived by Bikbaev \cite{Bikbaev:1985}, however, with insufficient 
care paid to details. (In fact, his terms $e^w$ and $e^{\hat w}$ 
on p.~581 are not
defined,  and in his formula (29), $(x,t)$-dependent 
terms are missing.)  More recently, algebro-geometric solutions of 
\eqref{T1.1} were also briefly considered by Wisse \cite{Wisse:1993}, 
again without explicitly deriving theta function representations for 
$u$ and $v$.

In Section \ref{Ts2} we follow Date's \cite{Date:1978} explicit 
realization of Mikhailov's commutator representation in terms of 
polynomials in the spectral parameter.  In Section \ref{Ts3} we 
develop the basic algebro-geometric formalism for \eqref{T1.2}, and 
from that point on we deviate from previous investigations and focus 
on a different approach based on the solution $\phi$ of a Riccati-type 
equation associated with the Thirring system \eqref{T1.2}.  We 
consider Dubrovin-type equations for auxiliary divisors and define 
the Baker--Akhiezer vector associated with the system \eqref{T1.2} in 
terms of the fundamental function $\phi$ on $\calK_{n}$, the 
underlying hyperelliptic curve of genus $n\in\bbN_{0}$. We also study
the algebro-geometric initial value problem in detail. Our principal 
results, the theta function representations of
$u$,
$v$,
$u^{*}$, 
$v^{*}$, and $\phi$ are derived in detail in Section \ref{Ts4}.  
Finally, Appendix \ref{A} collects some basic results on compact Riemann 
surfaces and introduces the terminology freely used in Sections 
\ref{Ts3} and \ref{Ts4}.

\section{The basic polynomial setup} \label{Ts2}

In this section we start from Mikhailov's \cite{Mikhailov:1976} 
commutator representation of the classical massive Thirring system in  
a form used by Date \cite{Date:1978} (see also 
\cite{HolodPrikarpatsky:1978},  \cite{PrikarpatskiiGolod:1979}, which 
contain similar material) in his analysis of quasi periodic solutions 
of this model.

Assuming $u, v, u^{*}, v^{*}\colon\bbR^2\to\bbC$ to satisfy 
\begin{align}
&u(\dott,t), u^{*}(\dott,t)\in C^1(\bbR), \quad 
v(\dott,t), v^{*}(\dott,t)\in C^{\infty}(\bbR), \quad t\in\bbR, \no \\
&u(x,\dott), u^{*}(x,\dott)\in C^1(\bbR), \quad 
\partial_x^k v(x,\dott), \partial_x^k v^{*}(x,\dott)\in C^1(\bbR),
\,\,\, k\in\bbN_0, \,\, x\in\bbR, \lb{T2.1}
\end{align}
we introduce the $2\times 2$ matrices
\begin{align} 
U(\zeta,x,t)&=i\begin{pmatrix}z-v(x,t)v^{*}(x,t) & 2\zeta v(x,t) \\
2\zeta v^{*}(x,t) & -z+v(x,t)v^{*}(x,t)\end{pmatrix}, \lb{T2.2} \\
V_{n+1}(\zeta,x,t)&=i
\begin{pmatrix}-G_{n+1}(z,x,t)& \zeta F_{n}(z,x,t)\\
    -\zeta H_{n}(z,x,t) &G_{n+1}(z,x,t)
\end{pmatrix}, \quad n\in\bbN_0, \lb{T2.3} \\
\ti V(\zeta,x,t)&=i
\begin{pmatrix}z^{-1}-u(x,t)u^{*}(x,t) & 2\zeta^{-1} u(x,t) \\
2\zeta^{-1} u^{*}(x,t) & -z^{-1}+u(x,t)u^{*}(x,t)\end{pmatrix}, 
 \lb{T2.4} \\
& \qquad\qquad\qquad\qquad \zeta\in\bbC\setminus\{0\}, \, z=\zeta^2, 
\, (x,t)\in\bbR^2, \no
\end{align}
where $F_{n}(z,x,t)$, $H_{n}(z,x,t)$, and $G_{n+1}(z,x,t)$ are 
polynomials with respect to $z$ of degree $n$ and $n+1$, 
respectively, that is, they are of the type
\begin{align} 
F_{n}(z,x,t)&=\sum_{j=0}^n 
f_{n-j}(x,t)z^j=f_{0}(x,t)\prod_{j=1}^n(z-\mu_{j}(x,t)), \lb{T2.5} \\
G_{n+1}(z,x,t)&=\sum_{j=0}^{n+1} 
g_{n+1-j}(x,t)z^j, \quad g_{0}(x,t)=1, \lb{T2.7} \\
H_{n}(z,x,t)&=\sum_{j=0}^n 
h_{n-j}(x,t)z^j=h_{0}(x,t)\prod_{j=1}^n(z-\nu_{j}(x,t)). \lb{T2.6} 
\end{align}
The classical massive Thirring system is then defined by demanding the 
zero-curvature representation
\begin{align} 
-V_{n+1,x}(\zeta,x,t)+[U(\zeta,x,t),V_{n+1}(\zeta,x,t)]&=0, \quad 
(\zeta,x,t)\in\bbC\setminus\{0\}\times\bbR^2, \lb{T2.8} \\
-V_{n+1,t}(\zeta,x,t)+[\ti V(\zeta,x,t),V_{n+1}(\zeta,x,t)]&=0, \quad 
(\zeta,x,t)\in\bbC\setminus\{0\}\times\bbR^2. \lb{T2.9}
\end{align}
Explicitly, equations \eqref{T2.8} and \eqref{T2.9} yield
\begin{align}
F_{n,x}(z,x,t)&=-2i(v(x,t)v^{*}(x,t)-z)F_{n}(z,x,t)+4iv(x,t)G_{n+1}(z,x,t),
\lb{T2.10} \\
G_{n+1,x}(z,x,t)&=2iz v^{*}(x,t)F_{n}(z,x,t)+2iz v(x,t)H_{n}(z,x,t),
\lb{T2.12} \\
H_{n,x}(z,x,t)&=2i(v(x,t)v^{*}(x,t)-z)H_{n}(z,x,t)+4iv^{*}(x,t)G_{n+1}(z,x,t),
\lb{T2.11} \\
\begin{split}
F_{n,t}(z,x,t)&=-2i(u(x,t)u^{*}(x,t)-z^{-1})F_{n}(z,x,t) \\
&\quad+4iz^{-1} u(x,t)G_{n+1}(z,x,t), \lb{T2.13} 
\end{split}\\
G_{n+1,t}(z,x,t)&=2i u^{*}(x,t)F_{n}(z,x,t)+2iu(x,t)H_{n}(z,x,t),
\lb{T2.15} \\
\begin{split}
H_{n,t}(z,x,t)&=2i(u(x,t)u^{*}(x,t)-z^{-1})H_{n}(z,x,t)\\
&\quad +4iz^{-1}u^{*}(x,t)G_{n+1}(z,x,t). \lb{T2.14} 
\end{split} 
\end{align} 
By \eqref{T2.10}--\eqref{T2.15} one infers that 
\begin{equation}
    \left(G_{n+1}^2-z F_{n}H_{n}  \right)_{x}
    =\left(G_{n+1}^2-z F_{n}H_{n}  \right)_{t}=0\lb{T2.16}
\end{equation} 
and hence
\begin{equation}
    G_{n+1}(z,x,t)^2-z F_{n}(z,x,t)H_{n}(z,x,t) =R_{2n+2}(z),\lb{T2.17}
\end{equation} 
where the integration constant $R_{2n+2}(z)$ is a monic polynomial 
in $z$ of degree $2n+2$, that is,
\begin{equation}
R_{2n+2}(z)=\prod_{m=0}^{2n+1}(z-E_{m}), \quad
\{E_{m}\}_{m=0,\dots,2n+1}\subset\bbC, \lb{T2.18}
\end{equation} 
since we chose $g_{0}=1$.  Moreover, \eqref{T2.17} implies
\begin{equation}
g_{n+1}(x,t)^2=\prod_{m=0}^{2n+1} E_{m}\lb{T2.19}
\end{equation} 
and we will choose
\begin{equation}
g_{n+1}\neq 0, \text{  that is, $E_{m}\neq 0$, 
$m=0,\dots,2n+1$}.\lb{T2.20}
\end{equation}
The actual sign of $g_{n+1}$ will be determined later (cf.
\eqref{T3.8}, \eqref{T3.9}).  A comparison  of coefficients of $z^k$
in \eqref{T2.10}--\eqref{T2.14} then yields
\begin{align}
f_0&=-2v, \no \\
f_1&=iv_x+2v^2v^{*}+c_1(-2v), \no \\
f_n&=-2g_{n+1}u, \no \\
g_0&=1, \no \\
g_1&=-2vv^{*}+c_1, \lb{T2.21} \\
g_{n+1}&=\bigg(\prod_{m=0}^{2n+1} E_m\bigg)^{1/2}, \no \\
h_0&=2v^{*}, \no \\
h_1&=iv^{*}_x-2v(v^{*})^2+c_1 2v^{*}, \no \\
h_n&=2g_{n+1}u^{*}, \text{ etc.,} \no
\end{align} 
where $\{c_\ell\}_{\ell\in\bbN}\subset\bbC$ denote integration
constants, and
\begin{align}
-iu_{x}(x,t)+2v(x,t)+2v(x,t)v^{*}(x,t)u(x,t)&=0, \lb{T2.22} \\
iu_{x}^{*}(x,t)+2v^{*}(x,t)+2v(x,t)v^{*}(x,t)u^{*}(x,t)&=0,
\lb{T2.23} \\ 
-iv_{t}(x,t)+2u(x,t)+2u(x,t)u^{*}(x,t)v(x,t)&=0,
\lb{T2.24} \\
iv_{t}^{*}(x,t)+2u^{*}(x,t)+2u(x,t)u^{*}(x,t)v^{*}(x,t)&=0. \lb{T2.25}
\end{align}

Equations \eqref{T2.22}--\eqref{T2.25} represent the classical massive 
Thirring system in light cone coordinates.  It should be emphasized 
that the original Thirring equations are given by \eqref{T2.22}, 
\eqref{T2.24} imposing the constraints
\begin{equation}
u^{*}(x,t)=\overline{u(x,t)}, \quad 
v^{*}(x,t)=\overline{v(x,t)}, \lb{T2.26}
\end{equation}
where the bar denotes the operation of complex conjugate.  In this paper,
however, we  will not impose the constraints \eqref{T2.26} but rather
study the  system \eqref{T2.22}--\eqref{T2.25}.

Given \eqref{T2.22}--\eqref{T2.25}, a straightforward computation
verifies the commutator relation
\begin{equation}
U_t(\zeta,x,t)-\ti V_x(\zeta,x,t) +[U(\zeta,x,t),\ti V(\zeta,x,t)]=0, 
\quad (\zeta,x,t)\in\bbC\setminus \{0\}\times \bbR^2, \lb{T2.26a}
\end{equation}
complementing \eqref{T2.8} and \eqref{T2.9}.

This concludes our brief review of the polynomial setup by Date 
 \cite{Date:1978}, and for the remainder of this paper we 
will deviate from his strategy and focus on an approach based on the 
solution $\phi$ of a Riccati-type equation associated with the 
Thirring system.  This will enable us to employ a formalism previously 
applied to the KdV, AKNS, Toda, Boussinesq, and the combined 
sine-Gordon--mKdV hierarchies \cite{BullaGesztesyHoldenTeschl:1997}, 
\cite{DicksonGesztesyUnterkofler:2000}, 
\cite{DicksonGesztesyUnterkofler1:2000}, \cite{GesztesyHolden2:1997}, 
\cite{GesztesyHolden:1999a}, \cite{GesztesyRatnaseelan:1996}, 
\cite{GesztesyRatnaseelanTeschl:1996}.

We conclude this section by mentioning the elementary fact that the 
Thirring system \eqref{T2.22}--\eqref{T2.25} is invariant under the 
scaling transformation,
\begin{equation}
(u, v, u^{*}, v^{*})\to (Au, Av, A^{-1}u^{*}, A^{-1}v^{*}), 
\quad A\in\bbC\setminus\{0\}. \lb{T2.27}
\end{equation} 
In the special case where $u^{*}=\uu, v^{*}=\vv$, $A$ in
\eqref{T2.27} is further constrained by $|A|=1$.

\section{The basic algebro-geometric formalism} \lb{Ts3}

Introducing the (possibly singular) hyperelliptic curve $\calK_{n}$ 
of (arithmetic) genus $n\in\bbN_{0}$, 
\begin{align}
    \calK_{n}\colon \calF_{n}(z,y)&=y^2-R_{2n+2}(z)=0, \lb{T3.1a} \\
    R_{2n+2}(z)&=\prod_{m=0}^{2n+1}(z-E_{m}), \quad 
    \{E_{m}\}_{m=0,\dots,2n+1}\subset\bbC\setminus\{0\}, \lb{T3.1}
\end{align}
we denote points $P$ on $\calK_{n}$ by $P=(z,y)$ and compactify 
$\calK_{n}$ by joining two points at infinity $\Pinfp$, 
$\Pinfm$, $\Pinfp\neq \Pinfm$, still denoting 
the compactified curve by $\calK_{n}$.  Moreover, we recall the 
hyperelliptic involution (sheet exchange map) $*$ on $\calK_{n}$,
\begin{equation} 
*\colon\calK_{n}\to\calK_{n}, \quad P=(z,y)\mapsto P^{*}=(z,-y), \, 
\Pinfp^{*}=\Pinfm.\lb{T3.2}
\end{equation} 
For additional facts on $\calK_{n}$ and further notation freely 
employed throughout this paper, the reader may want to consult 
Appendix \ref{A}.

Next, we define the fundamental meromorphic function 
$\phi(\dott,x,t)$ on $\calK_{n}$ by
\begin{align}
\phi(P,x,t)&=\frac{y(P)+G_{n+1}(z,x,t)}{F_{n}(z,x,t)}  \lb{T3.3} \\
&=\frac{-zH_{n}(z,x,t)}{y(P)-G_{n+1}(z,x,t)}, \quad 
P=(z,y)\in\calK_{n}, \, (x,t)\in\bbR^2,\lb{T3.4}
\end{align}
where we used \eqref{T2.17} to obtain \eqref{T3.4}.  Introducing
\begin{align} 
\hat\mu_{j}(x,t)&=(\mu_{j}(x,t),G_{n+1}(\mu_{j}(x,t),x,t))\in\calK_{n}, \quad
j=1,\dots,n, \, (x,t)\in\bbR^2, \lb{T3.5} \\
\hat\nu_{j}(x,t)&=(\nu_{j}(x,t),-G_{n+1}(\nu_{j}(x,t),x,t))\in\calK_{n}, \quad
j=1,\dots,n, \, (x,t)\in\bbR^2, \lb{T3.6}
\end{align}
and 
\begin{equation}
    \Pzeropm=(0,\pm G_{n+1}(0))=(0,\pm g_{n+1})\in\calK_{n}, \lb{T3.7}
\end{equation}
we fix the branch of $y(P)$ near $\Pinfpm$ according to
\begin{equation}
\lim_{\abs{z}\to\infty}\frac{y(P)}{G_{n+1}(z,x,t)}
=\lim_{\abs{z}\to\infty}\frac{y(P)}{z^{n+1}}=\mp 1
\text{  as $P\to \Pinfpm$}
\lb{T3.8}
\end{equation}
and consequently determine the sign of $g_{n+1}$, 
\begin{equation}
g_{n+1}=\left(\prod_{m=0}^{2n+1} E_{m} \right)^{1/2},
\lb{T3.9}
\end{equation}
by compatibility of all local charts on $\calK_{n}$. We note that  
$\Pzeropm$ and $\Pinfpm$ are not necessarily on the same sheet of 
$\calK_{n}$. The actual sheet on which $\Pzeropm$ lie depends on the 
sign of $g_{n+1}$ and hence on the location of all $E_{m}$. 

Given these 
conventions, the divisor $(\phi(\dott,x,t))$ of $\phi(\dott,x,t)$ then 
reads
\begin{equation}
(\phi(\dott,x,t))=\calD_{\Pzerom\hunu(x,t)}
-\calD_{\Pinfm\humu(x,t)}, \quad (x,t)\in\bbR^2.
\lb{T3.10}
\end{equation}
Next we collect a few characteristic properties of $\phi$.
\begin{lemma} \lb{lemmaT3.1} 
Assume \eqref{T2.1}, \eqref{T2.8}, \eqref{T2.9}, and \eqref{T3.1} and 
let $P=(z,y)\in\calK_{n}$, $(x,t)\in\bbR^2$.  Then $\phi$ satisfies 
the Riccati-type equations
\begin{align}
\begin{split}
   \phi_{x}(P,x,t)&+2iv(x,t)\phi(P,x,t)^2\\
&\qquad+2i(z-v(x,t)v^{*}(x,t))\phi(P,x,t)
  =2izv^{*}(x,t), \lb{T3.11} 
\end{split}\\
\begin{split}
   \phi_{t}(P,x,t)&+2iz^{-1}u(x,t)\phi(P,x,t)^2\\
   &\qquad+2i(z^{-1}-u(x,t)u^{*}(x,t))\phi(P,x,t)
  =2iu^{*}(x,t).\lb{T3.12}
\end{split}
\end{align}
Moreover,
\begin{align}
\phi(P,x,t)\phi(P^{*},x,t)&=zH_{n}(z,x,t)/F_{n}(z,x,t),\lb{T3.13} \\
\phi(P,x,t)+\phi(P^{*},x,t)&=2G_{n+1}(z,x,t)/F_{n}(z,x,t),\lb{T3.14} \\
\phi(P,x,t)-\phi(P^{*},x,t)&=2y(P)/F_{n}(z,x,t).\lb{T3.15}
\end{align}
\end{lemma}
\begin{proof}  Equation \eqref{T3.11} follows from 
\eqref{T2.10}--\eqref{T2.12}, \eqref{T2.17}, and \eqref{T3.3}. 
Similarly, \eqref{T3.12} follows from \eqref{T2.13}--\eqref{T2.15},
\eqref{T2.17}, and 
\eqref{T3.3}. Relations \eqref{T3.13}--\eqref{T3.15} are obvious from 
\eqref{T2.17} and \eqref{T3.3}.
\end{proof}

Given $\phi(P,x,t)$, we can define the Baker--Akhiezer vector 
$\Psi(P,\zeta,x,x_{0},t,t_{0})$ by
\begin{align}
\begin{split}
\Psi(P,\zeta,x,x_{0},t,t_{0})&=
\begin{pmatrix} \psi_{1}(P,x,x_{0},t,t_{0}) \\
\psi_{2}(P,\zeta,x,x_{0},t,t_{0})\end{pmatrix}, \\
&\quad P=(z,y)\in\calK_{n}\setminus\{\Pinfpm\}, \, z=\zeta^2, 
\, (x,t), (x_{0},t_{0})\in\bbR^2,  
\end{split}\lb{T3.16}
\end{align}
\begin{align}
\begin{split}
\psi_{1}(&P,x,x_{0},t,t_{0}) \\
&=\exp\bigg(i \int_{t_{0}}^t ds\,\big(z^{-1}-u(x_{0},s)u^{*}(x_{0},s) 
+2z^{-1}u(x_{0},s)\phi(P,x_{0},s)  \big)\\ 
&\quad+i\int_{x_{0}}^x dx'\,\big(z-v(x',t)v^{*}(x',t) 
+2v(x',t)\phi(P,x',t)  \big) \bigg), 
\end{split}\lb{T3.17} \\
\psi_{2}(&P,\zeta,x,x_{0},t,t_{0})=\zeta^{-1}\psi_{1}(P,x,x_{0},t,t_{0})
\phi(P,x,t). \lb{T3.18}
\end{align}
Properties of  $\Psi$ are summarized in the following result.
\begin{lemma}  \lb{lemmaT3.2}
Assume \eqref{T2.1}, \eqref{T2.8}, \eqref{T2.9}, and \eqref{T3.1} and 
let $P=(z,y)\in\calK_{n}\setminus\{\Pinfpm\}$, $(x,t), 
(x_{0},t_{0})\in\bbR^2$.  Then $\Psi(P,\zeta,x,x_{0},t,t_{0})$ 
satisfies
\begin{align}
\Psi_{x}(P,\zeta,x,x_{0},t,t_{0})&=
U(\zeta,x,t)\Psi(P,\zeta,x,x_{0},t,t_{0}), 
\lb{T3.19} \\
\Psi_{t}(P,\zeta,x,x_{0},t,t_{0})&=\ti V(\zeta,x,t)
\Psi(P,\zeta,x,x_{0},t,t_{0}), \lb{T3.20} \\
iy(P)\Psi(P,\zeta,x,x_{0},t,t_{0})&=V_{n+1}(\zeta,x,t)
\Psi(P,\zeta,x,x_{0},t,t_{0}). \lb{T3.20a} 
\end{align}
Moreover, if the zeros of $F_{n}(\dott,x,t)$ are all simple for 
$(x,t)\in\Omega$, $\Omega\subseteq\bbR^2$ open and connected, then 
$\psi_{1}(\dott,x,x_{0},t,t_{0})$, $(x,t), (x_{0},t_{0})\in\Omega$,  
is meromorphic on $\calK_{n}\setminus\{\Pinfpm\}$.  In addition,
\begin{multline}
\psi_{1}(P,x,x_{0},t,t_{0})=
\left(\f{F_{n}(z,x,t)}{F_{n}(z,x_{0},t_{0})} \right)^{1/2} \times \\
\quad \times\exp\bigg(2i y(P)z^{-1}\int_{t_{0}}^t 
ds\,\f{u(x_{0},s)}{F_{n}(z,x_{0},s)}+2i y(P)\int_{x_{0}}^x 
dx'\,\f{v(x',t)}{F_{n}(z,x',t)}   \bigg), \lb{T3.21} 
\end{multline}
\begin{align}
\psi_{1}(P,x,x_{0},t,t_{0})\psi_{1}(P^{*},x,x_{0},t,t_{0})&=
F_{n}(z,x,t)/F_{n}(z,x_{0},t_{0}), \lb{T3.22}  \\
\psi_{2}(P,\zeta,x,x_{0},t,t_{0})\psi_{2}(P^{*},\zeta,x,x_{0},t,t_{0})&=
H_{n}(z,x,t)/F_{n}(z,x_{0},t_{0}), \lb{T3.23}  \\
\psi_{1}(P,x,x_{0},t,t_{0})\psi_{2}(P^{*},\zeta,x,x_{0},t,t_{0})
+&\psi_{1}(P^{*},x,x_{0},t,t_{0})\psi_{2}(P,\zeta,x,x_{0},t,t_{0}) \no
\\
&=2\zeta^{-1} G_{n+1}(z,x,t)/F_{n}(z,x_{0},t_{0}). \lb{T3.24} 
\end{align}
\end{lemma}
\begin{proof} Equations \eqref{T3.19}, \eqref{T3.20} are verified 
using \eqref{T2.10}--\eqref{T2.15}, \eqref{T3.11}, \eqref{T3.12}, 
\eqref{T3.17}, and \eqref{T3.18}. \eqref{T3.20a} follows by
combining \eqref{T2.3}, \eqref{T3.3}, \eqref{T3.4},
\eqref{T3.17}, and \eqref{T3.18}. Clearly $\psi_{1}$ is 
 meromorphic on 
$\calK_{n}\setminus\{\Pinfpm,\hmu_{1}(x,t),\dots,\hmu_{n}(x,t) 
\}$ by \eqref{T3.17}. Since 
\begin{align}
2iv(x',t)\phi(P,x',t)&\underset{P\to\hmu_{j}(x',t)}{=}
\f{\partial}{\partial x'}\ln\big(F_{n}(z,x',t)  \big)+\Oh(1)
\text{  as $z\to\mu_{j}(x',t)$}, \lb{T3.25} \\ 
2iz^{-1}u(x_{0},s)\phi(P,x_{0},s)&\underset{P\to\hmu_{j}(x_{0},s)}{=}
\f{\partial}{\partial s}\ln\big(F_{n}(z,x_{0},s)  \big)+\Oh(1)
\text{  as $z\to\mu_{j}(x_{0},s)$}, \lb{T3.26} 
\end{align}
one infers that $\psi_{1}$ is meromorphic on 
$\calK_{n}\setminus\{\Pinfpm\}$ if the zeros of $F_{n}(\dott,x,t)$ are all 
simple. This follows from \eqref{T3.17} by restricting $P$ to
a sufficiently small neighborhood $\calU_j(x_{0})$ of
$\{\hmu_j(x_{0},s)\in\calK_{n}\,|\,(x_{0},s)\in\Omega, \,s\in
[t_{0},t] \}$ such that $\hmu_k(x_{0},s)\notin \calU_j(x_{0})$
for all $s\in [t_{0},t]$ and all
$k\in\{1,\dots,n\}\setminus\{j\}$, and similarly, by restricting 
$P$ to a sufficiently small neighborhood $\calU_j(t)$ of
$\{\hmu_j(x^{\prime},t)\in\calK_{n}\,|\,(x^{\prime},t)\in\Omega,
\,x^{\prime}\in [x_{0},x] \}$ such that $\hmu_k(x^{\prime},t)\notin
\calU_j(t)$ for all $x^{\prime}\in [x_{0},x]$ and all
$k\in\{1,\dots,n\}\setminus\{j\}$. Equation
\eqref{T3.21}  follows from \eqref{T3.17} after replacing $\phi$ by 
the right-hand 
side of \eqref{T3.3} and utilizing \eqref{T2.10} in the $x'$-integral 
and \eqref{T2.13} in the $s$-integral.  Equations 
\eqref{T3.22}--\eqref{T3.24} immediately follow from 
\eqref{T3.13}--\eqref{T3.15}, and \eqref{T3.18}.
\end{proof}
Next we discuss the asymptotic behavior of $\phi(P,x,t)$ as $P\to 
\Pzeropm, \Pinfpm$ in some detail since this will turn out 
to be a crucial ingredient for the theta function representation to be 
derived in Section \ref{Ts4}.

\begin{lemma}\lb{lemmaT3.3}
Assume \eqref{T2.1}, \eqref{T2.8}, \eqref{T2.9}, and \eqref{T3.1}. 
Then
\begin{align}
\phi(P,x,t)&\underset{z\to\infty}{=}-\f{1}{v(x,t)}z
+\f{i}{2}\bigg(\f{1}{v(x,t)} \bigg)_{x}+\Oh\bigg(\f{1}{z}\bigg) 
\text{  as 
$P=(z,y)\to \Pinfm$}, \lb{T3.27} \\
\phi(P,x,t)&\underset{z\to\infty}{=}v^{*}(x,t)
+\f{i}{2}v^{*}_{x}(x,t) \f{1}{z}+\Oh\bigg(\f{1}{z^2}\bigg) \text{  as 
$P=(z,y)\to \Pinfp$}, \lb{T3.28} \\
\phi(P,x,t)&\underset{z\to 0}{=}u^{*}(x,t)z
+\f{i}{2}u^{*}_{t}(x,t) z^2+\Oh(z^3) \text{  as 
$P=(z,y)\to \Pzerom$}, \lb{T3.29} \\
\phi(P,x,t)&\underset{z\to 0}{=}-\f{1}{u(x,t)}
+\f{i}{2}\bigg(\f{1}{u(x,t)}\bigg)_{t}z+\Oh(z^2) \text{  as 
$P=(z,y)\to \Pzerop$}. \lb{T3.30} 
\end{align}
\end{lemma}
\begin{proof} The existence of these asymptotic expansions (in terms 
of local coordinates $\zeta=1/z$ near $\Pinfpm$ and local 
coordinate $\zeta=z$ near $\Pzeropm$) is clear from the explicit 
form of $\phi$ in \eqref{T3.3}.  Insertion of the polynomials 
$F_{n}$, $H_{n}$, and $G_{n+1}$, then in principle, yields the explicit 
expansion coefficients in \eqref{T3.27}--\eqref{T3.30}.  However, 
this is a cumbersome procedure, especially with regard to the next to 
leading coefficients in \eqref{T3.27}--\eqref{T3.30}. Much more 
efficient is the actual computation of these coefficients utilizing 
the Riccati-type equations \eqref{T3.11} and \eqref{T3.12}.  Indeed, 
inserting the ansatz
\begin{equation}
    \phi\underset{z\to\infty}{=}z\phi_{-1}+\phi_{0}
+\Oh\bigg(\f{1}{z}\bigg)  \lb{T3.31}
\end{equation}
into \eqref{T3.11} and comparing the first two leading powers of $z$ 
immediately yields \eqref{T3.27}.  Similarly, the ansatz
\begin{equation}
     \phi\underset{z\to\infty}{=}\phi_{0}+\phi_{1}\f{1}{z}
+\Oh\bigg(\f{1}{z^2}\bigg) \lb{T3.32}
\end{equation}
inserted into \eqref{T3.11} immediately produces \eqref{T3.28}.  In 
exactly the same manner, inserting the ansatz
\begin{equation}
     \phi\underset{z\to 0}{=}\phi_{1}z+\phi_{2}z^2+\Oh(z^3) 
    \lb{T3.33}
\end{equation}
and the ansatz
\begin{equation}
     \phi\underset{z\to 0}{=}\phi_{0}+\phi_{1}z+\Oh(z^2) 
    \lb{T3.34}
\end{equation}
into \eqref{T3.12} immediately yields \eqref{T3.29} and 
\eqref{T3.30}, respectively.
\end{proof} 

We follow up with a similar asymptotic analysis of 
$\psi_{1}(P,x,x_{0},t,t_{0})$.
\begin{lemma} \lb{lemmaT3.4}
    Assume \eqref{T2.1}, \eqref{T2.8}, \eqref{T2.9}, and 
    \eqref{T3.1}.  Then
\begin{align}
\psi_{1}(P,x,x_{0},t,t_{0})&\underset{z\to\infty}{=}
\exp\big(\mp iz(x-x_{0})+\Oh(1)  \big) \text{  as $P=(z,y)\to 
\Pinfmp$}, \lb{T3.34a} \\
\psi_{1}(P,x,x_{0},t,t_{0})&\underset{z\to 0}{=}
\exp\big(\pm iz^{-1}(t-t_{0})+\Oh(1)  \big) \text{  as $P=(z,y)\to 
\Pzeromp$}. \lb{T3.34b}
\end{align}
\end{lemma}
\begin{proof} Equations \eqref{T3.34a} and \eqref{T3.34b} follow 
from \eqref{T3.17} noting
\begin{align}
\begin{split}
i(z-v(x,t)v^{*}(x,t))+2iv(x,t)\phi(&P,x,t)\underset{z\to\infty}{=}
\mp i z+\Oh(1)\\
&  \text{  as $P=(z,y)\to 
\Pinfmp$}, 
\end{split} \lb{T3.34c}\\
\begin{split}
i(z^{-1}-u(x_{0},s)u^{*}(x_{0},s))+2iz^{-1}u(x_{0},&s)\phi(P,x_{0},s)
\underset{z\to\infty}{=}
\Oh(1) \\
& \text{  as $P=(z,y)\to \Pinfmp$},  
\end{split}\lb{T3.34d}\\
\begin{split}
i(z-v(x,t)v^{*}(x,t))+2iv(&x,t)\phi(P,x,t)\underset{z\to 0}{=}
\Oh(1)\\
& \text{  as $P=(z,y)\to \Pzeromp$}, 
\end{split}\lb{T3.34e} \\
\begin{split}
i(z^{-1}-u(x_{0},s)u^{*}(x_{0},s))+2iz^{-1}u(x_{0},s)\phi(P,x_{0}&,s)
\underset{z\to 0}{=}
\pm i z^{-1}+\Oh(1)\\
&\text{  as $P=(z,y)\to \Pzeromp$}. \end{split}\lb{T3.34f}
\end{align}
\end{proof}

In some of the following considerations it is appropriate to assume
that $\calK_{n}$ is nonsingular and hence we then assume 
\begin{equation}
E_m \neq E_{m^\prime} \text{ for } m \neq m^\prime, \,\, 
m, m^\prime=0,\dots,2n+1 \lb{T3.34g}
\end{equation}
in addition to \eqref{T3.1}.

Next, we turn to Dubrovin-type equations for 
$\mu_{j}(x,t)$, $\nu_{j}(x,t)$, $j=1,\dots,n$, that is, we derive the 
nonlinear first-order system of partial differential equations 
governing their $(x,t)$-variation.

\begin{lemma} \lb{lemmaT3.5}
 Let $n\in\bbN$.~Assume \eqref{T2.1}, \eqref{T2.8}, \eqref{T2.9},  
 and \eqref{T3.1} and suppose that the zeros
$\{\mu_{j}(x,t)\}_{j=1,\dots,n}$ of  $F_{n}(\dott,x,t)$ remain
distinct for $(x,t)\in\ti\Omega_{\mu}$, where 
    $\ti\Omega_{\mu}\subseteq\bbR^2$ is open and connected.  Then 
    $\{\mu_{j}(x,t)\}_{j=1,\dots,n}$ satisfies the following system 
    of differential equations
\begin{align}
\mu_{j,x}(x,t)&=2iy(\hmu_{j}(x,t))\prod_{\substack{ \ell=1\\ 
\ell\neq j}}^n (\mu_{j}(x,t)-\mu_{\ell}(x,t))^{-1}, \lb{T3.35} \\
\mu_{j,t}(x,t)&=(-1)^ng_{n+1}^{-1}
\bigg(\prod_{\substack{ k=1\\ k\neq j}}^n \mu_{k}(x,t)\bigg) 
2iy(\hmu_{j}(x,t))\prod_{\substack{ \ell=1\\ 
\ell\neq j}}^n (\mu_{j}(x,t)-\mu_{\ell}(x,t))^{-1}, \no \\
& \hspace*{5.5cm} j=1,\dots,n, \, (x,t)\in\ti\Omega_{\mu}. \lb{T3.36} 
\end{align}
Next, assume $\calK_{n}$ to be nonsingular and introduce the
initial condition
\begin{equation}  
\{\hmu_{j}(x_{0},t_{0})\}_{j=1,\dots,n}\subset\calK_{n}, \lb{T3.37} 
\end{equation}
where $\{\mu_{j}(x_0,t_0)\}_{j=1,\dots,n}$ remain distinct and
distinct from zero. Then there exists an open and connected set 
$\Omega_{\mu}\subseteq\bbR^2$, with $(x_0,t_0)\in\Omega_{\mu}$, 
such that the initial value problem 
\eqref{T3.35}--\eqref{T3.37} has a unique solution 
$\{\hmu_{j}(x,t)\}_{j=1,\dots,n}$ satisfying
\begin{equation}
\hmu_{j}\in C^\infty(\Omega_{\mu},\calK_{n}), \quad j=1,\dots,n.\lb{T3.38} 
\end{equation}
For the zeros $\{\nu_{j}(x,t)\}_{j=1,\dots,n}$ of  $H_{n}(\dott,x,t)$ 
identical statements hold with $\mu$ replaced by $\nu$,
$\ti\Omega_{\mu}$ by $\ti\Omega_{\nu}$, etc.  In  particular,
$\{\hnu_{j}(x,t)\}_{j=1,\dots,n}$ satisfies
\begin{align}
\nu_{j,x}(x,t)&=2iy(\hnu_{j}(x,t))\prod_{\substack{ \ell=1\\ 
\ell\neq j}}^n (\nu_{j}(x,t)-\nu_{\ell}(x,t))^{-1}, \lb{T3.39} \\
\nu_{j,t}(x,t)&=(-1)^ng_{n+1}^{-1}
\bigg(\prod_{\substack{ k=1\\ k\neq j}}^n \nu_{k}(x,t)\bigg) 
2iy(\hnu_{j}(x,t))\prod_{\substack{ \ell=1\\ 
\ell\neq j}}^n (\nu_{j}(x,t)-\nu_{\ell}(x,t))^{-1}, \no \\
& \hspace*{5.5cm} j=1,\dots,n, \, (x,t)\in\ti\Omega_{\nu}. \lb{T3.40} 
\end{align}
\end{lemma}
\begin{proof}  Equations \eqref{T2.5}, \eqref{T2.10}, and 
\eqref{T3.5} imply
\begin{align}
    F_{n,x}(\mu_{j})=f_{0}(-\mu_{j,x})\prod_{\substack{ 
    \ell=1\\\ell\neq j}}^n (\mu_{j}-\mu_{\ell})
    =4iv G_{n+1}(\mu_{j})=4ivy(\hmu_{j}). \lb{T3.41}
\end{align}
Using $f_{0}=-2v$ by \eqref{T2.21}, one concludes \eqref{T3.35}.  
Similarly, one derives from \eqref{T2.5}, \eqref{T2.13}, and 
\eqref{T3.5},
\begin{equation}
F_{n,t}(\mu_{j})=f_{0}(-\mu_{j,t})\prod_{\substack{ \ell=1 \\ \ell\neq 
j}}^n (\mu_{j} -\mu_{\ell})=(4iu/\mu_{j}) G_{n+1}(\mu_{j})
=(4iu/\mu_{j}) y(\hmu_{j}). \lb{T3.42}
\end{equation}
Since
\begin{equation}
-4iu/f_{0}=2if_{n}/(f_{0}g_{n+1})
=2i(-1)^n\big(\prod_{k=1}^n\mu_{k}\big)/g_{n+1}    \lb{T3.43}
\end{equation}
by \eqref{T2.21} and \eqref{T2.5}, one arrives at \eqref{T3.36}.  
Equations \eqref{T3.39} and \eqref{T3.40} are derived analogously. In
order to conclude \eqref{T3.38}, one first needs to investigate the
case where $\hmu_{j}(x,t)$ hits one of the branch points
$(E_m,0)\in\calB(\calK_{n})$ and hence the right-hand sides of
\eqref{T3.35} and \eqref{T3.36} vanish. Thus we suppose that
\begin{equation}
\mu_{j_0}(x,t)\to E_{m_0} \text{ as } (x,t)\to 
(\tilde x_0,\tilde t_0) \lb{T3.43a}
\end{equation} 
for some $j_0\in\{1,\dots,n\}$, $m_0\in\{0,\dots,2n+1\}$ and some 
$(\tilde x_0,\tilde t_0)\in\Omega_\mu$. Introducing
\begin{equation}
\zeta_{j_0}(x,t)=(\mu_{j_0}(x,t)-E_{m_0})^{1/2}, \quad 
\mu_{j_0}(x,t)=E_{m_0}+\zeta_{j_0}(x,t)^2 \lb{T3.43b}
\end{equation}
for $(x,t)$ in an open neighborhood of
$(\tilde x_0,\tilde t_0)\in\Omega_{\mu}$, equations 
\eqref{T3.35} and \eqref{T3.36} become
\begin{align}
\zeta_{j_0,x}(x,t)\underset{(x,t)\to(\tilde
x_0,\tilde t_0)}{=}&2i\bigg(
\prod_{\substack{ m=0 \\ m\neq
m_0}}^{2n+1}\big(E_{m_0}-E_m\big)\bigg)^{1/2}\bigg(
\prod_{\substack{ k=1 \\ k\neq
j_0}}^{n}\big(E_{m_0}-\mu_k(x,t)\big)^{-1}\bigg)\times \no \\
&\times(1+\Oh(\zeta_{j_0}(x,t)^2)), \lb{T3.43c} \\
\zeta_{j_0,t}(x,t)\underset{(x,t)\to(\tilde
x_0,\tilde t_0)}{=}&2i\bigg(
\prod_{\substack{ m=0 \\ m\neq
m_0}}^{2n+1}\big(E_{m_0}-E_m\big)\bigg)^{1/2}
\bigg(\prod_{\substack{ k=1 \\ k\neq
j_0}}^{n}\big(E_{m_0}-\mu_k(x,t)\big)^{-1}\bigg)\times \no \\
&\times\bigg(\prod_{\substack{ \ell=1 \\ \ell\neq
j_0}}^{n}\mu_{\ell}(x,t)\bigg)(1+\Oh(\zeta_{j_0}(x,t)^2)).  
\lb{T3.43d}
\end{align}
Since by hypothesis the right-hand sides of \eqref{T3.43c} and
\eqref{T3.43d} are nonvanishing, one arrives at \eqref{T3.38}.
\end{proof}

Next we derive a few trace formulas involving $u, v, u^{*}, v^{*}$
and some of their $x$-derivatives in terms of $\mu_{j}(x,t)$ and 
$\nu_{j}(x,t)$.

\begin{lemma} \lb{lemmaT3.6}
 Let $n\in\bbN$ and assume \eqref{T2.1}, \eqref{T2.8}, \eqref{T2.9},  
 and \eqref{T3.1}. Then
\begin{align}
&i\f{v_x(x,t)}{v(x,t)}+2v(x,t)v^{*}(x,t)-2c_1
=2\sum_{j=1}^n \mu_j(x,t), \lb{T3.44} \\
&i\f{v_x(x,t)}{v(x,t)}-2v(x,t)v^{*}(x,t)
=-i\sum_{j=1}^n \f{\mu_{j,x}(x,t)}{\mu_j(x,t)} + 
\f{2(-1)^n g_{n+1}}{\prod_{j=1}^n \mu_j(x,t)}, \lb{T3.45} \\
&\f{v(x,t)}{u(x,t)}=\f{(-1)^n g_{n+1}}{\prod_{j=1}^n \mu_j(x,t)},
\lb{T3.46} \\
&i\f{v_x^{*}(x,t)}{v^{*}(x,t)}-2v(x,t)v^{*}(x,t)+2c_1=-2\sum_{j=1}^n
\nu_j(x,t), 
\lb{T3.47} \\
&i\f{v_x^{*}(x,t)}{v^{*}(x,t)}+2v(x,t)v^{*}(x,t)=-i\sum_{j=1}^n
\f{\nu_{j,x}(x,t)}{\nu_j(x,t)}-\f{2(-1)^ng_{n+1}}{\prod_{j=1}^n
\nu_j(x,t)}, \lb{T3.48} \\
&\f{v^{*}(x,t)}{u^{*}(x,t)}=\f{(-1)^n g_{n+1}}{\prod_{j=1}^n
\nu_j(x,t)}.
\lb{T3.49}
\end{align} 
Here
\begin{equation}
c_1=-\f{1}{2}\sum_{m=0}^{2n+1} E_m \lb{T3.50} 
\end{equation}
and $g_{n+1}= \big(\prod_{m=0}^{2n+1} E_m\big)^{1/2}$ has been
introduced in \eqref{T3.8} and \eqref{T3.9}.
\end{lemma}
\begin{proof}
Equations \eqref{T3.44} and \eqref{T3.47} follow from \eqref{T2.5},  
\eqref{T2.6} by comparing powers of $z^n$ and $z^{n-1}$, using 
\eqref{T2.21}. \eqref{T3.45} and \eqref{T3.48} follow from taking 
$z=0$ in \eqref{T2.10} and \eqref{T2.11}, using again
\eqref{T2.21}. Finally, \eqref{T3.46} and \eqref{T3.49} follow from 
$f_n=f_0\prod_{j=1}^n(-\mu_j)$, $h_n=h_0\prod_{j=1}^n(-\nu_j)$ and 
\eqref{T2.21}. 
\end{proof}

While we are not explicitly introducing the hierarchy of massive 
Thirring equations in this paper, we note that Dubrovin-type equations 
such as \eqref{T3.35}, \eqref{T3.36} combined with trace formulas for 
$u, v, u^{*}, v^{*}$ in terms of $\mu_{j}(x,t)$, enable one to
discuss  such a hierarchy following the approach outlined in 
\cite{GesztesyHolden:1999c}.

Up to this point we assumed the zero curvature equations 
\eqref{T2.8} and
\eqref{T2.9}, or  equivalently, \eqref{T2.10}--\eqref{T2.14} and as a
consequence, derived the corresponding algebro-geometric formalism. 
In the 
remainder of this section we will study the algebro-geometric initial 
value problem, that is, starting from the Dubrovin equations 
\eqref{T3.35}--\eqref{T3.37} and the trace formulas 
\eqref{T3.44}--\eqref{T3.46}, derive 
\eqref{T2.10}--\eqref{T2.14}, and hence the zero curvature equations 
\eqref{T2.8} and \eqref{T2.9}.

We start with an elementary result extending the scaling
transformation mentioned in \eqref{T3.27}. 

\begin{lemma} \lb{lemmaT3.7}
Assume \eqref{T2.1} and suppose $u, v, u^{*}, v^{*}$ 
satisfy the Thirring system \eqref{T2.22}--\eqref{T2.25}.
Assume $A(t)=\exp\big(\int^t ds\,a(s)\big)$, $t\in\bbR$, with $a\in
C(\bbR)$ and consider the time-dependent scaling
transformation
\begin{equation}
(u, v, u^{*}, v^{*})\to (\breve{u}, \breve{v}, \breve{u}^{*}, 
\breve{v}^{*})=(Au, Av, A^{-1}u^{*}, A^{-1}v^{*}). \lb{T3.52}
\end{equation} 
Then $\breve{u}, \breve{v}, \breve{u}^{*}, \breve{v}^{*}$ satisfy 
the corresponding extended massive Thirring system 
\begin{align}
-i\breve{u}_{x}(x,t)+2\breve{v}(x,t)+
2\breve{v}(x,t)\breve{v}^{*}(x,t)\breve{u}(x,t)&=0, \lb{T3.53} \\
i\breve{u}_{x}^{*}(x,t)+2\breve{v}^{*}(x,t)+
2\breve{v}(x,t)\breve{v}^{*}(x,t)\breve{u}^{*}(x,t)&=0, \lb{T3.54} \\
-i\breve{v}_{t}(x,t)+2\breve{u}(x,t)+
2\breve{u}(x,t)\breve{u}^{*}(x,t)\breve{v}(x,t)
+ia(t)\breve{v}(x,t)&=0, \lb{T3.55} \\
i\breve{v}_{t}^{*}(x,t)+2\breve{u}^{*}(x,t)+
2\breve{u}(x,t)\breve{u}^{*}(x,t)\breve{v}^{*}(x,t)-
ia(t)\breve{v}^{*}(x,t)&=0. \lb{T3.56}
\end{align}
\end{lemma}
\begin{proof}
It suffices to insert \eqref{T3.52} into the system 
\eqref{T2.22}--\eqref{T2.25}.
\end{proof}
In the special case where $u^{*}(x,t)=\ol{u(x,t)},
v^{*}(x,t)=\ol{v(x,t)}$, $A(t)$ in Lemma \ref{lemmaT3.7} is further
constrained by $|A(t)|=1$, $t\in\bbR$.

Next we provide the basic setup for the algebro-geometric initial
value problem. We start from the following assumptions. 

\begin{hypothesis} \lb{hypothesisT3.8} 
Given the hyperelliptic curve
$\calK_{n}$ in \eqref{T3.1}, and the proper choice of the branch of
$g_{n+1}$ defined by $g_{n+1}=\big(\prod_{m=0}^{2n+1}E_m\big)^{1/2}$,
according to
\eqref{T3.8} {\rm (}i.e., according to $\lim_{|z|\to\infty}
y(P)z^{-n-1}=\mp\infty$ as $P\to\Pinfpm${\rm )}, consider the
Dubrovin-type system of differential equations \eqref{T3.35},
\eqref{T3.36} on $\Omega_{\mu}$, for some intitial conditions
\eqref{T3.37}. Here
$\Omega_{\mu}\subseteq\bbR^2$ is assumed to be open and connected, and
such that the projections
$\mu_j(x,t)$ of $\hmu_j(x,t)$ onto $\bbC$ remain distinct and
distinct from zero for $(x,t)\in\Omega_{\mu}$, that is,
\begin{align}
&\mu_j(x,t)\ne\mu_{j^{\prime}}(x,t) \text{ for } j\ne j^{\prime}, 
\,\, j,j^{\prime}=1,\dots,n, \,\, (x,t)\in\Omega_{\mu}, 
\lb{T3.57} \\
&\{\mu_j(x,t)\}_{j=1,\dots,n}\cap\{0\}
=\emptyset, \quad (x,t)\in\Omega_{\mu}. \lb{T3.58}
\end{align}
\end{hypothesis}

Assuming Hypothesis \ref{hypothesisT3.8} in the following, we will
next define $u,v,u^{*},v^{*}$ and the polynomials $F_{n}, G_{n+1},
H_{n}$ in the following steps (S1)--(S4).\\

\noindent (S1). Use the trace formulas
\eqref{T3.44}--\eqref{T3.46} on $\Omega_{\mu}$, that is, 
\begin{align}
&i\f{v_x(x,t)}{v(x,t)}+2v(x,t)v^{*}(x,t)-2c_1
=2\sum_{j=1}^n \mu_j(x,t), \lb{T3.63} \\
&i\f{v_x(x,t)}{v(x,t)}-2v(x,t)v^{*}(x,t)
=-i\sum_{j=1}^n \f{\mu_{j,x}(x,t)}{\mu_j(x,t)} + 
\f{2(-1)^n g_{n+1}}{\prod_{j=1}^n
\mu_j(x,t)}, \lb{T3.64} \\ 
&u(x,t)=(-1)^n g_{n+1}^{-1}v(x,t)\prod_{j=1}^n \mu_j(x,t),
\quad (x,t)\in\Omega_{\mu}, \lb{T3.65}
\end{align}
to define $u(x,t), v(x,t), v^{*}(x,t)$ on $\Omega_{\mu}$ up to a
possibly
$t$-dependent multiple factor according to the scale transformation
described in  Lemma \ref{lemmaT3.7}.\\

\noindent (S2). Define the polynomial $F_{n}(z,x,t)$ on
$\bbc\times\Omega_{\mu}$ of degree $n$ with respect to $z$ by
\begin{equation}
F_{n}(z,x,t)=-2v(x,t)\prod_{j=1}^n(z-\mu_j(x,t)), \quad 
(z,x,t)\in\bbC\times\Omega_{\mu} \lb{T3.66}  
\end{equation}
and define the polynomial $G_{n+1}(z,x,t)$ on
$\bbc\times\Omega_{\mu}$ of degree $n+1$ with respect to $z$ by
\begin{align}
&F_{n,x}(z,x,t)=-2i(v(x,t)v^{*}(x,t)-z)F_{n}(z,x,t)+
4iv(x,t)G_{n+1}(z,x,t), \lb{T3.67} \\
& \hspace*{8.2cm} (z,x,t)\in\bbC\times\Omega_{\mu}. \no
\end{align}
One then verifies from
\begin{equation}
2iy(\hmu_j)=\mu_{j,x} \prod_{\substack{k=1\\ 
k\neq j}}^n (\mu_j-\mu_k)=\f{F_{n,x}(\mu_j)}{2v}, 
\quad j=1,\dots,n \lb{T3.68}
\end{equation}
and \eqref{T3.67} that 
\begin{equation}
y(\hmu_j(x,t))=\f{F_{n,x}(\mu_j(x,t)),x,t)}{4iv(x,t)}
=G_{n+1}(\mu_j(x,t),x,t), \quad j=1,\dots,n, \,\,
(x,t)\in\Omega_{\mu} \lb{T3.69}
\end{equation}
and hence
\begin{equation}
\big(G_{n+1}(z,x,t)^2-R_{2n+2}(z)\big)\big|_{z=\mu_j(x,t)}=0, \quad 
j=1,\dots,n, \,\, (x,t)\in\Omega_{\mu}. \lb{T3.70}
\end{equation}

\vspace*{2mm}

\noindent (S3). Taking $z=0$ in \eqref{T3.67}, using
\eqref{T3.66}, results in
\begin{equation}
\f{2(-1)^nG_{n+1}(0)}{\prod_{j=1}^n\mu_j}=i\f{v_x}{v}-2vv^{*}+
i\sum_{j=1}^n\f{\mu_{j,x}}{\mu_j} \lb{T3.71}
\end{equation}
and hence a comparison with \eqref{T3.64} yields
\begin{equation}
G_{n+1}(0,x,t)=g_{n+1}=\bigg(\prod_{m=0}^{2n+1}E_m\bigg)^{1/2}
\lb{T3.72}
\end{equation}
and thus, 
\begin{equation}
\big(G_{n+1}(z,x,t)^2-R_{2n+2}(z)\big)\big|_{z=0}=0, \quad 
 (x,t)\in\Omega_{\mu}. \lb{T3.73}
\end{equation}
Because of \eqref{T3.70} and \eqref{T3.73} we can define a
polynomial $H_{n}(z,x,t)$ on $\bbC\times\Omega_{\mu}$ of degree $n$
with respect to $z$ by
\begin{equation}
G_{n+1}(z,x,t)^2-R_{2n+2}(z)=zF_{n}(z,x,t)H_{n}(z,x,t), \quad 
(z,x,t)\in\bbC\times\Omega_{\mu}. \lb{T3.74}
\end{equation}

\vspace*{2mm}

\noindent (S4). Given $H_{n}(z,x,t)$ we finally define $u^{*}(x,t)$
on $\Omega_{\mu}$ by
\begin{equation}
u^{*}(x,t)=\f{H_{n}(0,x,t)}{2g_{n+1}}, 
\quad  (x,t)\in\Omega_{\mu}, \lb{T3.75}
\end{equation}
Again $u^{*}(x,t)$ is defined up to a possibly $t$-dependent factor
in accordance with Lemma \ref{lemmaT3.7}.

The algebro-geometric initial value problem now can be solved as
follows.

\begin{theorem}\lb{theoremT3.9}
Assume Hypothesis \ref{hypothesisT3.8}, define $u,v,u^{*},v^{*}$ and 
 $F_{n}, G_{n+1}, H_{n}$ as in $(S1)-(S4)$ and let
$(x,t)\in\Omega_{\mu}$. Then there exists a function $a\in
C^\infty(\Omega_{\mu})$, independent of 
$x$ {\rm (}$a_x|_{\Omega_{\mu}}=0${\rm )}, such that 
\begin{align}
F_{n,x}(z,x,t)&=-2i(v(x,t)v^{*}(x,t)-z)F_{n}(z,x,t)+
4iv(x,t)G_{n+1}(z,x,t),
\lb{T3.76} \\
G_{n+1,x}(z,x,t)&=2iz v^{*}(x,t)F_{n}(z,x,t)+2iz v(x,t)H_{n}(z,x,t),
\lb{T3.78} \\
H_{n,x}(z,x,t)&=2i(v(x,t)v^{*}(x,t)-z)H_{n}(z,x,t)+
4iv^{*}(x,t)G_{n+1}(z,x,t),
\lb{T3.77} \\
\begin{split}
F_{n,t}(z,x,t)&=-2i(u(x,t)u^{*}(x,t)-z^{-1})F_{n}(z,x,t) 
+a(t)F_{n}(z,x,t)\\
&\,\quad+4iz^{-1} u(x,t)G_{n+1}(z,x,t), \lb{T3.79} 
\end{split}\\
G_{n+1,t}(z,x,t)&=2i u^{*}(x,t)F_{n}(z,x,t)+2iu(x,t)H_{n}(z,x,t),
\lb{T3.81} \\
\begin{split}
H_{n,t}(z,x,t)&=2i(u(x,t)u^{*}(x,t)-z^{-1})H_{n}(z,x,t)
-a(t)H_{n}(z,x,t)\\
&\,\quad +4iz^{-1}u^{*}(x,t)G_{n+1}(z,x,t). \lb{T3.80} 
\end{split}
\end{align}
In particular, $u, v, u^{*}, v^{*}$ satisfy the
extended massive Thirring system \eqref{T3.53}--\eqref{T3.56} on
$\Omega_{\mu}$, 
\begin{align}
-i{u}_{x}(x,t)+2{v}(x,t)+2{v}(x,t){v}^{*}(x,t){u}(x,t)&=0, 
\lb{T3.82} \\ 
i{u}_{x}^{*}(x,t)+2{v}^{*}(x,t)+2{v}(x,t){v}^{*}(x,t){u}^{*}(x,t)&=0,
\lb{T3.83} \\ 
-i{v}_{t}(x,t)+2{u}(x,t)+2{u}(x,t){u}^{*}(x,t){v}(x,t)
+ia(t){v}(x,t)&=0, \lb{T3.84} \\
i{v}_{t}^{*}(x,t)+2{u}^{*}(x,t)+2{u}(x,t){u}^{*}(x,t){v}^{*}(x,t)-
ia(t){v}^{*}(x,t)&=0. \lb{T3.85}
\end{align}
\end{theorem}
\begin{proof}
Define the polynomial
\begin{align}
&P_{n}(z,x,t)=2izv^{*}(x,t)F_{n}(z,x,t)+2izv(x,t)H_{n}(z,x,t)-
G_{n+1,x}(z,x,t),  \no  \\
& \hspace*{8cm} (z,x,t)\in\bbC\times\Omega_{\mu}. \lb{T3.86}
\end{align}
Using \eqref{T3.69} and
$2G_{n+1}G_{n+1,x}=z(F_{n,x}H_{n}+F_{n}H_{n,x})$ (by
differentiating \eqref{T3.74} with respect to $x$) one then computes
\begin{align}
G_{n+1}(\mu_j)P_{n}(\mu_j)&=2i\mu_jvH_{n}(\mu_j)G_{n+1}(\mu_j)-
G_{n+1}(\mu_j)G_{n+1,x}(\mu_j) \no \\
&=\f12\mu_jH_{n}(\mu_j)F_{n,x}(\mu_j)-
\f12\mu_jF_{n,x}(\mu_j)H_{n}(\mu_j)=0,  \lb{T3.87} \\
&\hspace*{5.8cm}  j=1,\dots,n. \no
\end{align}
In order to investigate the leading-order term with respect to
$z$ of $P_{n}(z)$ we first study the leading-order $z$-behavior
of $F_{n}(z), G_{n+1}(z)$, and $H_{n}(z)$. Writing (cf. 
\eqref{T2.5}--\eqref{T2.6})
\begin{align}
F_{n}(z)=\sum_{j=0}^n f_{n-j}z^j, \quad
H_{n}(z)=\sum_{j=0}^n h_{n-j}z^j,  \quad
G_{n+1}(z)=\sum_{j=0}^{n+1}  g_{n+1-j}z^j, \,\,
g_{0}=1, \lb{T3.88}
\end{align}
a comparison of leading powers with respect to $z$ in \eqref{T3.66},
\eqref{T3.67}, and \eqref{T3.74} yields
\begin{align}
f_0&=-2v, \lb{T3.89}\\
g_0&=1, \lb{T3.90} \\
v_x+2iv^2v^{*}+if_1+2ig_1&=0, \lb{T3.91} \\
2g_1+2vh_0+\sum_{m=0}^{2n+1}E_m&=0. \lb{T3.92} 
\end{align}
Since \eqref{T3.63} can be rewritten in the form
\begin{equation}
f_1=iv_x+2v^2v^{*}+v\sum_{m=0}^{2n+1}E_m, \lb{T3.93}
\end{equation}
a comparison of \eqref{T3.91} and \eqref{T3.93} then yields
\begin{equation}
g_1=-2vv^{*}-\f12 \sum_{m=0}^{2n+1}E_m \lb{T3.94}
\end{equation}
and hence
\begin{equation}
h_0=2v^{*}. \lb{T3.95}
\end{equation}
Insertion of \eqref{T3.89}, \eqref{T3.90}, and \eqref{T3.95} into 
\eqref{T3.86} then yields
\begin{equation}
P_{n}(z,x,t)=\Oh(z^{n}) \text{ as } |z|\to\infty. \lb{T3.96}
\end{equation} 
Thus, \eqref{T3.87} and \eqref{T3.96} prove
\begin{equation}
P_{n}(z,x,t)=b(x,t)F_{n}(z,x,t), \quad
(z,x,t)\in\bbC\times\Omega_{\mu} \lb{T3.97}
\end{equation}
for some $b\in C^\infty(\Omega_\mu)$ (independent of $z$), implying
\begin{align}
&G_{n+1,x}(z,x,t)=2izv^{*}(x,t)F_{n}(z,x,t)+2izv(x,t)H_{n}(z,x,t)- 
b(x,t)F_n(z,x,t),  \no \\
& \hspace*{8.3cm} (z,x,t)\in\bbC\times\Omega_{\mu}. \lb{T3.98}
\end{align}
Taking $z=0$ in \eqref{T3.98}, observing that $G_{n+1}(0,x,t)$ is
independent of $(x,t)\in\Omega_{\mu}$ by \eqref{T3.72}, then shows
that
\begin{equation}
0=-b(x,t)F_n(0,x,t), \quad (x,t)\in\Omega_{\mu}, \lb{T3.99}
\end{equation}
and hence $b=0$ on $\Omega_{\mu}$ because of \eqref{T3.58}. Thus,
\begin{equation}
G_{n+1,x}(z,x,t)=2izv^{*}(x,t)F_{n}(z,x,t)+2izv(x,t)H_{n}(z,x,t), 
\quad (z,x,t)\in\bbC\times\Omega_{\mu}. \lb{T3.100}
\end{equation} 
Differentiating \eqref{T3.74} with respect to $x$, inserting 
\eqref{T3.67} and \eqref{T3.100}, then yields
\begin{align}
&H_{n,x}(z,x,t)=2i(v(x,t)v^{*}(x,t)-z)H_{n}(z,x,t)+
4iv^{*}(x,t)G_{n+1}(z,x,t), \lb{T3.101} \\
& \hspace*{8.3cm} (z,x,t)\in\bbC\times\Omega_{\mu} \no
\end{align}
and we proved \eqref{T3.76}--\eqref{T3.77}. 

Next, combining \eqref{T3.36}, \eqref{T3.65}, and \eqref{T3.69} one
computes
\begin{align}
F_{n,t}(\mu_j)&=2v\f{(-1)^n}{g_{n+1}}\bigg(\prod_{\substack{k=1\\ 
k\neq j}}^n \mu_k\bigg) 2iy(\hmu_j)=\f{(-1)^n}{g_{n+1}}
\bigg(\prod_{\substack{k=1\\ k\neq j}}^n \mu_k\bigg)\f{4iv}{\mu_j}
G_{n+1}(\mu_j) \no \\
&=\f{4iu}{\mu_j}G_{n+1}(\mu_j), \quad j=1,\dots,n. \lb{T3.102}
\end{align}
Since clearly 
\begin{equation}
F_{n,t}(z)-\big(-2i(uu^{*}-z^{-1})F_{n}(z)+4iz^{-1}uG_{n+1}(z)\big)
=\Oh(z^n) \text{ as } |z|\to\infty, \lb{T3.103}
\end{equation}
a comparison of \eqref{T3.102} and \eqref{T3.103} yields
\begin{align}
&F_{n,t}(z,x,t)-\big(-2i(u(x,t)u^{*}(x,t)-z^{-1})F_{n}(z,x,t))
+4iz^{-1}uG_{n+1}(z,x,t)\big) \no \\
&=a(x,t)F_{n}(z,x,t), \quad (z,x,t)\in\bbC\times\Omega_{\mu}
\lb{T3.104}
\end{align}
for some $a\in C^\infty(\Omega_{\mu})$ (independent of $z$), and
hence \eqref{T3.79} (except for $a_x=0$). A comparison of powers of
$z^n$ in \eqref{T3.104} then yields \eqref{T3.84}. 

Next, we restrict $\Omega_{\mu}$ a bit further and introduce 
$\ti\Omega_{\mu}\subseteq\Omega_{\mu}$ by the requirement that 
$\mu_j(x,t)$ remain distinct and also 
distinct from $\{E_m\}_{m=0,\dots,2n+1}\cup\{0\}$ for
$(x,t)\in\ti\Omega_{\mu}$, that is, we suppose
\begin{align}
&\mu_j(x,t)\ne\mu_{j^{\prime}}(x,t) \text{ for } j\ne j^{\prime}, 
\,\, j,j^{\prime}=1,\dots,n, \,\, (x,t)\in\ti\Omega_{\mu}, 
\lb{T3.105} \\
&\{\mu_j(x,t)\}_{j=1,\dots,n}\cap\{\{E_m\}_{m=0,\dots,2n+1}\cup\{0\}\}
=\emptyset, \quad (x,t)\in\ti\Omega_{\mu}. \lb{T3.105a}
\end{align}
Differentiating
\eqref{T3.74} with respect to $t$ inserting \eqref{T3.104} then yields
\begin{align}
2G_{n+1}(z)G_{n+1,t}(z)&=
zF_{n}(z)\big(-2i(uu^{*}-z^{-1})H_{n}(z)+aH_{n}(z)+H_{n,t}(z)\big) 
\no \\
&\quad \, +4iuG_{n+1}H_{n}(z). \lb{T3.106}
\end{align}
Since the zeros of $F_{n}$ and $G_{n+1}$ are disjoint by hypothesis
\eqref{T3.105a} (cf.~also \eqref{T3.69}), 
$zH_{n,t}(z)$ necessarily must be of the form
\begin{align}
zH_{n,t}(z,x,t)&=2i(zu(x,t)u^{*}(x,t)-1)H_{n}(z,x,t)
-a(x,t)zH_{n}(z,x,t) \no \\
&\quad \, +4id(x,t)G_{n+1}(z,x,t), \quad 
(z,x,t)\in\bbC\times\ti\Omega_{\mu} \lb{T3.107} 
\end{align}
for some $d\in C^\infty(\ti\Omega_{\mu})$ (independent of $z$) and
\eqref{T3.107} inserted into \eqref{T3.106} then yields 
\begin{equation}
G_{n+1}(z,x,t)=2iu(x,t)H_{n}(z,x,t)+2id(x,t)F_{n}(z,x,t), \quad 
(z,x,t)\in\bbC\times\ti\Omega_{\mu}. \lb{T3.108} 
\end{equation}
Since
\begin{equation}
u(x,t)=-\f{F_{n}(0,x,t)}{2g_{n+1}}, 
\quad  (x,t)\in\ti\Omega_{\mu}, \lb{T3.109}
\end{equation}
combining \eqref{T3.65} and \eqref{T3.66}, taking $z=0$ in
\eqref{T3.108}, observing \eqref{T3.72} and \eqref{T3.75}, results in
\begin{equation}
0=2iu2g_{n+1}u^{*}+2id(-2g_{n+1}u) \lb{T3.110}
\end{equation}
and hence in
\begin{equation}
d(x,t)=u^{*}(x,t), \quad (x,t)\in\ti\Omega_{\mu}. \lb{T3.111}
\end{equation}
Using property \eqref{T3.38}, \eqref{T3.107}--\eqref{T3.111} then
extend by continuity from $\ti\Omega_{\mu}$ to $\Omega_{\mu}$. This
proves
\eqref{T3.81} and \eqref{T3.80} (except for
$a_x=0$). A comparison of powers  of $z^n$ in \eqref{T3.80} then
yields \eqref{T3.85}. Taking
$z=0$ in
\eqref{T3.76} and \eqref{T3.77}, observing \eqref{T3.75} and
\eqref{T3.109}, then proves \eqref{T3.82} and \eqref{T3.83}. Finally,
computing the partial $t$-derivative of $F_{n,x}$ and separately the
partial $x$-derivative of $F_{n,t}$, utilizing \eqref{T3.76},
\eqref{T3.78}, \eqref{T3.79}, \eqref{T3.81}, and 
\eqref{T3.82}--\eqref{T3.85} then shows
\begin{equation}
F_{n,xt}(z,x,t)-F_{n,tx}(z,x,t)=-a_x(x,t)F_{n}(z,x,t), \quad
(z,x,t)\in\bbC\times\Omega_{\mu} \lb{T3.112}
\end{equation}
and hence 
\begin{equation}
a_x(x,t)=0, \quad (x,t)\in\Omega_{\mu}. \lb{T3.113}
\end{equation}
\end{proof}

\begin{remark} \lb{remarkT3.10}
(i) The fact that the system of Dubrovin equations
\eqref{T3.35}--\eqref{T3.37}  cannot uniquely determine the solutions
$u,v,u^{*},v^{*}$ of the massive Thirring system
\eqref{T2.22}--\eqref{T2.25}, as is evident from the occurrence of
$a(t)$ in \eqref{T3.84}, \eqref{T3.85}, is of course due to the scale
covariance displayed explicitly in Lemma
\ref{lemmaT3.7}. In particular, once a certain $a(t)$ has been
identified, a scaling transformation of the type \eqref{T3.52} (with
$A(t)$ replaced by $1/A(t)$) will restore the extended massive
Thirring system \eqref{T3.82}--\eqref{T3.85} to it's original form in 
\eqref{T2.22}--\eqref{T2.25}. \\ 
(ii) For simplicity we formulated Theorem \ref{theoremT3.9} in terms 
of $\{\hmu_j\}_{j=1,\dots,n}$ and \eqref{T3.35}--\eqref{T3.37} only.
Of course there exists a completely analogous approach starting with
$\{\hnu_j\}_{j=1,\dots,n}$ and the system \eqref{T3.39}, \eqref{T3.40}
instead. \\   
(iii) Invoking the explicit theta function representations for 
$u,v,u^{*},v^{*}$ to be proven in Section \ref{Ts4} next (this
approach is independent of that used to prove Theorem \ref{T3.9}),
one can extend the principal assertions
\eqref{T3.76}--\eqref{T3.85} of Theorem \ref{T3.9} by continuity
to $(x,t)$ lying in a larger set $\Omega\subseteq\bbR^2$ as long as
the divisors $\calD_{\humu(x,t)}$ and $\calD_{\hunu(x,t)}$ remain
nonspecial for $(x,t)\in\Omega$ (cf. Theorem \ref{theoremT4.3} and
Theorem \ref{taa20}).  
\end{remark}

\section{Theta function representations} \lb{Ts4}

In our final section we now derive theta function representations for 
the principal objects of Section \ref{Ts3}, including $\phi$, 
$\psi_{1}$, $u$, $v$, $u^{*}$, $v^{*}$.  These representations 
complement the papers by Date \cite{Date:1978} and Prikarpatskii 
and Golod 
\cite{PrikarpatskiiGolod:1979}, where theta function representations 
were derived for appropriate symmetric functions associated 
with auxiliary
divisors,  but not explicitly for $u$, $v$, $u^{*}$, $v^{*}$. 
Moreover, we correct some inaccuracies of such formulas in a 
paper by Bikbaev
\cite{Bikbaev:1985} (which follows a different strategy than ours).

According to our shift in emphasis from the Baker--Akhiezer vector 
$\Psi$ to our fundamental meromorphic function $\phi$ on $\calK_{n}$, 
we next aim at the theta function representation of $\phi$.

Assuming $\calK_{n}$ to be nonsingular for the remainder of this 
section (i.e., $E_{m}\neq E_{m'}$ for $m\neq m'$, 
$m,m'=0,\dots,2n+1$) and $n\in\bbN$ for simplicity (to avoid 
repeated case distinctions), we next recall the formula for
a normal  differential of the third kind, which has simple 
poles at $\Pzerom$ 
and $\Pinfm$, corresponding residues $+1$ and $-1$, 
vanishing $a$-periods, and is holomorphic otherwise on $\calK_{n}$.  
One computes
\begin{equation}
\omega_{\Pzerom, \Pinfm}^{(3)}=\f{y+y_{0,-}}{2z}\,\f{dz}{y}
+\f{\prod_{j=1}^n(z-\lambda_{j})dz}{2y}, \quad
\Pzerom=(0, y_{0,-})=(0,-g_{n+1}), \lb{T4.1}
\end{equation}
where $\{\lambda_{j}\}_{j=1,\dots,n}$ are uniquely determined by the 
normalization
\begin{equation}
\int_{a_{j}}\omega_{\Pzerom, \Pinfm}^{(3)}=0, \quad 
j=1,\dots,n.\lb{T4.2}
\end{equation}
The explicit formula \eqref{T4.1} then implies (using the local 
coordinate $\zeta=z$ near $\Pzeromp$)
\begin{equation}
\omega_{\Pzerom, \Pinfm}^{(3)}(P)\underset{\zeta\to 0}{=}
\left\{\begin{matrix}\zeta^{-1} \\ 0\end{matrix} \right\} \, d\zeta
\pm \left(\sum_{q=0}^\infty (q+1)\omega_{q+1}^0\zeta^q\right)d\zeta 
\text{  as $P\to \Pzeromp$}, \lb{T4.3}
\end{equation}
and similarly (using the local coordinate $\zeta=1/z$ near 
$\Pinfmp$),
\begin{equation}
\omega_{\Pzerom, \Pinfm}^{(3)}(P)\underset{\zeta\to 0}{=}
\left\{\begin{matrix} -\zeta^{-1} \\ 0\end{matrix} \right\} \, d\zeta
\pm \left(\sum_{q=0}^\infty (q+1)\omega_{q+1}^\infty\zeta^q\right)d\zeta 
\text{  as $P\to \Pinfmp$}. \lb{T4.4}
\end{equation}
In particular,
\begin{align}
\int_{Q_0}^P \omega_{\Pzerom, \Pinfm}^{(3)}&\underset{\zeta\to 0}{=}
\left\{\begin{matrix} \ln(\zeta) \\ 0\end{matrix} \right\}
+\omega_0^{0,\mp}
\pm \omega_{1}^0 \zeta \pm \omega_2^0 \zeta^2 + \Oh(\zeta^3) 
\text{  as $P\to \Pzeromp$}, \lb{T4.3a} \\
\int_{Q_0}^P \omega_{\Pzerom, \Pinfm}^{(3)}&\underset{\zeta\to 0}{=}
\left\{\begin{matrix} -\ln(\zeta) \\ 0\end{matrix} \right\}
+\omega^{\infty_\mp}_0
\pm \omega_1^\infty \zeta \pm \omega_2^\infty \zeta^2 +\Oh(\zeta^3) 
\text{  as $P\to \Pinfmp$}. \lb{T4.4a}
\end{align}
Here $Q_0\in\calB(\calK_{n})$ is an appropriate base point and we
agree to choose the same path of integration from $Q_0$ to $P$ in all
Abelian integrals in this section. 

A comparison of \eqref{T4.3}, \eqref{T4.4} with \eqref{T4.1}, 
\eqref{b27a}, and \eqref{b27ac} then yields
\begin{align}
\omega_{1}^0&=\f{1}{4}\sum_{m=0}^{2n+1} \f{1}{E_m} 
- \f{(-1)^n}{2g_{n+1}} \prod_{j=1}^n \lambda_j, \lb{T4.4b} \\
\omega_1^\infty&=-\f{1}{4}\sum_{m=0}^{2n+1} E_m 
+ \f{1}{2} \sum_{j=1}^n \lambda_j. \lb{T4.4c}
\end{align}
Next, we intend to go a step further and derive alternative 
expressions for the expansion coefficients $\omega_0^{0,\pm}$, 
$\omega_{1}^0$, $\omega^{\infty_\pm}_0$, and $\omega_1^\infty$ in 
\eqref{T4.3a} and \eqref{T4.4a}. To begin these calculations we
first recall the  notion of a nonsingular odd half-period $\Upsilon$
defined by
\begin{equation}
2\ul \Upsilon =0 \pmod{L_\N}, \quad \theta(\ul \Upsilon)=0, \,\, 
\frac{\partial \theta(\ul z)}{\partial z_j}\bigg|_{\ul z=\ul\Upsilon}
\neq 0 \text{ for some } j\in\{1,\dots,n\}. \lb{T4c}
\end{equation}
Discussions of even and odd half-periods (singular and nonsingular 
ones) can be found, for instance, in \cite[p.~12--15]{Fay:1973}, 
\cite{Lewittes:1964}. In addition, it is convenient to introduce 
the notation
\begin{align}
{\ul \Delta}_0&=\ua_{Q_{0}}(\Pzerop), \quad {\ul
\Delta}_\infty=\ua_{Q_{0}}(\Pinfp), \lb{T4d} \\
{\ul W}^0_1&=(W_{1,1}^0,\dots,W_{1,n}^0), \,\, 
W_{1,j}^0=\frac{1}{2\pi i} \int_{b_j} \omega^{(2)}_{\Pzerop,0}
=\frac{c_j (1)}{g_{n+1}}, \,\, j=1,\dots,n, \lb{T4e} \\
{\ul W}^0_2&=(W_{2,1}^0,\dots,W_{2,n}^0), \,\, 
W_{2,j}^0=\frac{c_j (1)}{4g_{n+1}}\sum_{m=0}^{2n+1}E_m^{-1}+
\f{c_j (2)}{2g_{n+1}}, \,\, j=1,\dots,n, \lb{T4ea} \\
{\ul W}_1^\infty&=(W^{\infty}_{1,1},\dots,W^{\infty}_{1,n}), \,\, 
W^{\infty}_{1,j}=\frac{1}{2\pi i} \int_{b_j} \omega^{(2)}_{\Pinfp,0}
=c_j (n), \,\, j=1,\dots,n, \lb{T4f} \\
{\ul W}_2^\infty&=(W^{\infty}_{2,1},\dots,W^{\infty}_{2,n}), \,\, 
W^{\infty}_{2,j}=\f{c_j (n)}{4}\sum_{m=0}^{2n+1}E_m+
\f{c_j (n-1)}{2}, \,\, j=1,\dots,n. \lb{T4fb}
\end{align}
Moreover, we abbreviate directional derivatives of $f$ in the 
direction of $\ul W=(W_1,\dots,W_n)\in\bbC^n$ by
\begin{align}
&(\partial_{\ul W}f)(\ul z)=\sum_{j=1}^n W_j\frac{\partial f}
{\partial z_j}(\ul z), \quad (\partial_{\ul W}^2f)(\ul z)=
\sum_{j,k=1}^n W_jW_k\frac{\partial^2 f}{\partial z_j \partial
z_k}(\ul z), \lb{T4fa} \\
& \hspace*{6.55cm} \ul z=(z_1,\dots,z_n)\in\bbC^n. \no
\end{align}
Then one obtains the following result.
\begin{lemma} \lb{lemmaT4.1a}
Given \eqref{T4.1}--\eqref{T4f} one obtains
\begin{align}
\omega_0^{0,+}&=\ln\bigg(\frac{\theta(\ul\Upsilon-2\ul\Delta_0)
\theta(\ul\Upsilon-\ul\Delta_\infty)}
{\theta(\ul\Upsilon-\ul\Delta_0-\ul\Delta_\infty)
\theta(\ul\Upsilon-\ul\Delta_0)}\bigg), \lb{T4g} \\
\omega_0^{0,-}&=\ln\bigg(\frac{(\partial_{\ul W^0_1}\theta)
(\ul\Upsilon)\theta(\ul\Upsilon-\ul\Delta_\infty)}
{\theta(\ul\Upsilon+\ul\Delta_0-\ul\Delta_\infty)
\theta(\ul\Upsilon-\ul\Delta_0)}\bigg), \lb{T4h} \\
\omega_1^0&=-\partial_{\ul W^0_1}\ln(\theta(\ul \Upsilon+
\ul\Delta_0-\ul\Delta_\infty))+
\f{(\partial_{\ul W^0_2}\theta)(\ul\Upsilon)+2^{-1}
(\partial^2_{\ul W^0_1}\theta)(\ul\Upsilon)}{(\partial_{\ul 
W^0_1}\theta)(\ul\Upsilon)} \lb{T4i}
\\ &=\partial_{\ul W^0_1}\ln\bigg(\frac{\theta(\ul\Upsilon
-2\ul\Delta_0)}{\theta(\ul\Upsilon-\ul\Delta_0-\ul\Delta_\infty)}
\bigg), \lb{T4j} \\
\omega_0^{\infty_+}&=\ln\bigg(\frac{\theta(\ul\Upsilon-\ul\Delta_0
-\ul\Delta_\infty)\theta(\ul\Upsilon-\ul\Delta_\infty}
{\theta(\ul\Upsilon-2\ul\Delta_\infty)\theta(\ul\Upsilon
-\ul\Delta_0)}\bigg), \lb{T4k} \\
\omega_0^{\infty_-}&=-\ln\bigg(\frac{(\partial_{\ul 
W^\infty_1}\theta)(\ul\Upsilon)\theta(\ul\Upsilon-\ul\Delta_0)}
{\theta(\ul\Upsilon-\ul\Delta_0+\ul\Delta_\infty)
\theta(\ul\Upsilon-\ul\Delta_\infty)}\bigg), \lb{T4l} \\
\omega_1^\infty&=\partial_{\ul W^\infty_1}\ln(\theta(\ul\Upsilon
-\ul\Delta_0+\ul\Delta_\infty))-\f{(\partial_{\ul 
W^\infty_2}\theta)(\ul\Upsilon)+2^{-1}(\partial^2_{\ul 
W^\infty_1}\theta)(\ul\Upsilon)}{(\partial_{\ul 
W^\infty_1}\theta)(\ul\Upsilon)} \lb{T4m}
\\ &=\partial_{\ul W^\infty_1}\ln\bigg(
\frac{\theta(\ul\Upsilon-\ul\Delta_0-\ul\Delta_\infty)}{\theta(
\ul\Upsilon-2\ul\Delta_\infty)}\bigg). \lb{T4n}
\end{align}   
\end{lemma}
\begin{proof}
Abbreviating
\begin{equation}
\ul w (P,Q_0)=\ul\Upsilon-\ua_{Q_{0}}(P)+\ua_{Q_{0}}(Q) 
\pmod{L_\N}, \lb{T4fo}
\end{equation}
one infers from 
\begin{align}
\ua_{Q_{0}}(P)&\underset{\zeta\to 
0}{=}\ua_{Q_{0}}(\Pzeropm)\pm \ul W_1^0\zeta \pm
\ul W_2^0\zeta^2+\Oh(\zeta^3) \text{  as $P\to \Pzeropm$}, 
\lb{T4.13} \\
\ua_{Q_{0}}(P)&\underset{\zeta\to 
0}{=}\ua_{Q_{0}}(\Pinfpm)\pm\ul W_1^\infty\zeta \pm \ul
W_2^\infty\zeta^2+\Oh(\zeta^3)\text{  as $P\to\Pinfpm$} \lb{T4.14} 
\end{align}
(cf.\ \eqref{b27} and \eqref{b27ab}), and \eqref{T4e},  
\eqref{T4f}, that 
\begin{align}
\theta(\ul w(P,Q))\underset{\zeta\to 0}{=}&\theta(\ul w(\Pzeropm,Q))
\mp (\partial_{\ul W_1^0}\theta)(\ul w(\Pzeropm,Q))\zeta 
\mp (\partial_{\ul W_2^0}\theta)(\ul w(\Pzeropm,Q))\zeta^2 \no \\
&+2^{-1} (\partial^2_{\ul W_1^0}\theta)(\ul w(\Pzeropm,Q))\zeta^2
+\Oh(\zeta^3) 
\text{ as $P\to\Pzeropm$}, \lb{T4p} \\
\theta(\ul w(P,Q))\underset{\zeta\to 0}{=}&\theta(\ul w(\Pinfpm,Q)) 
\mp (\partial_{\ul W_1^\infty}\theta)(\ul w(\Pinfpm,Q))\zeta 
\mp (\partial_{\ul W_2^\infty}\theta)(\ul w(\Pzeropm,Q))\zeta^2 \no \\
&+2^{-1} (\partial^2_{\ul W_1^\infty}\theta)(\ul w(\Pzeropm,Q))\zeta^2
+\Oh(\zeta^3) \text{ as $P\to\Pinfpm$}. \lb{T4q}
\end{align}
Next, observing the fact that
\begin{equation}
\omega^{(3)}_{\Pzerom,\Pinfm}=d\,\log\bigg(\frac{\theta(\ul
w(\dott,\Pzerom))}{\theta(\ul w(\dott,\Pinfm))}\bigg), \lb{T4r}
\end{equation}
it becomes a straightforward matter deriving \eqref{T4g}--\eqref{T4n}. 
For simplicity we just focus on the expansion of 
$\int_{Q_0}^P\omega^{(3)}_{\Pzerom,\Pinfm}$ as $P\to\Pzeropm$, 
the rest is completely analogous. Using 
\begin{align}
\ul w(Q_{0},\Pzeropm)&=\ul\Upsilon \pm\ul\Delta_0, \quad
\ul w(Q_{0},\Pinfpm)=\ul\Upsilon \pm\ul\Delta_\infty, 
\quad \ul w(Q,Q)=\ul\Upsilon, \,\, Q\in\calK_{n}, \no \\
\ul w(P_{\infty_\sigma},P_{0,\sigma^\prime})&=
\ul\Upsilon + \sigma^\prime \ul\Delta_0- 
\sigma \ul\Delta_\infty, \quad 
\ul w(P_{0,\sigma^\prime},P_{\infty_\sigma})=\ul\Upsilon 
- \sigma^\prime \ul\Delta_0+ \sigma \ul\Delta_\infty , \no \\
\ul w(P_{0,\sigma},P_{0,\sigma^\prime})&=
\ul\Upsilon+(\sigma^\prime
-\sigma)\ul\Delta_0, \quad \ul
w(P_{0,\sigma^\prime},P_{0,\sigma})=\ul\Upsilon+(\sigma 
-\sigma^\prime)\ul\Delta_0, \no \\ 
\ul w(P_{\infty_\sigma},P_{\infty_{\sigma^\prime}})&=
\ul\Upsilon+(\sigma^\prime-\sigma)\ul\Delta_\infty, 
\quad \ul
w(P_{\infty_{\sigma^\prime}},P_{\infty_\sigma})
=\ul\Upsilon+(\sigma-\sigma^\prime)\ul\Delta_\infty, 
\lb{T4ra} \\ 
&\hspace*{6.43cm} \sigma, \sigma^\prime \in \{1,-1\}, \no
\end{align}
and \eqref{T4p}--\eqref{T4r}, one computes by comparison with
\eqref{T4.3a},
\begin{align}
&\int_{Q_0}^P\omega^{(3)}_{\Pzerom,\Pinfm} = \int_{Q_0}^P
d\,\log\bigg(\frac{\theta(\ul w(P^\prime,\Pzerom))} {\theta(\ul
w(P^\prime,\Pinfm))}\bigg) \no \\ 
&=\ln\bigg(\frac{\theta(\ul w(P,\Pzerom))}
{\theta(\ul w(P,\Pinfm))}\bigg)-
\ln\bigg(\frac{\theta(\ul w(Q_{0},\Pzerom))}
{\theta(\ul w(Q_{0},\Pinfm))}\bigg) \no \\
&=\ln\bigg(\frac{\theta(\ul w(P,\Pzerom))}
{\theta(\ul w(P,\Pinfm))}\bigg)-
\ln\bigg(\frac{\theta(\ul\Upsilon-\ul\Delta_0)}
{\theta(\ul\Upsilon-\ul\Delta_\infty)}\bigg) \no \\
&=\ln\bigg(\frac{\theta(\ul\Upsilon-2\ul\Delta_0)
\theta(\ul\Upsilon-\ul\Delta_\infty)}
{\theta(\ul\Upsilon-\ul\Delta_0-\ul\Delta_\infty)
\theta(\ul\Upsilon-\ul\Delta_0)}\bigg)-
\partial_{\ul W^0_1}\ln\bigg(\frac{\theta(\ul\Upsilon
-2\ul\Delta_0)}{\theta(\ul\Upsilon-\ul\Delta_0-\ul\Delta_\infty)}
\bigg)\zeta+\Oh(\zeta^2) \no \\
&=\omega_0^{0,+}-\omega_1^0\zeta+\Oh(\zeta^2) 
\text{ as $P\to\Pzerop$}. \lb{T4s}
\end{align}
This proves \eqref{T4g} and \eqref{T4j}. Similarly, one calculates,
\begin{align}
&\int_{Q_0}^P\omega^{(3)}_{\Pzerom,\Pinfm} 
=\ln\bigg(\frac{\theta(\ul w(P,\Pzerom))}
{\theta(\ul w(P,\Pinfm))}\bigg)-
\ln\bigg(\frac{\theta(\ul\Upsilon-\ul\Delta_0)}
{\theta(\ul\Upsilon-\ul\Delta_\infty)}\bigg) \no \\
&=\ln(\zeta)+\ln\bigg(\frac{(\partial_{\ul W^0_1}\theta)(\ul\Upsilon)
\theta(\ul\Upsilon-\ul\Delta_\infty)}
{\theta(\ul\Upsilon+\ul\Delta_0-\ul\Delta_\infty)
\theta(\ul\Upsilon-\ul\Delta_0)}\bigg)-
\partial_{\ul W^0_1}\ln(\theta
(\ul\Upsilon+\ul\Delta_0-\ul\Delta_\infty))\zeta \no \\
& \quad + 
\f{(\partial_{\ul W^0_2}\theta)(\ul\Upsilon)+2^{-1}(\partial^2_{\ul
W^0_1}\theta)(\ul\Upsilon)}{(\partial_{\ul
W^0_1}\theta)(\ul\Upsilon)}\zeta +\Oh(\zeta^2) \no\\  
&=\ln(\zeta)+\omega_0^{0,-}+\omega_1^0\zeta+\Oh(\zeta^2) 
\text{ as $P\to\Pzerom$}, \lb{T4t}
\end{align}
proving \eqref{T4h} and \eqref{T4i}.
\end{proof}

The results of Lemma \ref{lemmaT4.1a} can conveniently be reformulated
in terms of theta functions with characteristics associated with the
vector $\ul\Upsilon$, but we omit further details at this point.

Combining \eqref{T3.10} and Theorem~\ref{taa17a}, the theta function 
representation of $\phi$ must be of the form
\begin{multline}
\phi(P,x,t)=C(x,t)\f{\theta\big(\uxi_{Q_{0}}-\ua_{Q_{0}}(P)
+\ual_{Q_{0}}(\calD_{\hunu(x,t)})\big)}
{\theta\big(\uxi_{Q_{0}}-\ua_{Q_{0}}(P)
+\ual_{Q_{0}}(\calD_{\humu(x,t)})\big)}\exp\left(\int_{Q_{0}}^P 
\omega_{\Pzerom,\Pinfm}^{(3)}\right), \\
P\in\calK_{n}, \, (x,t)\in\Omega, \lb{T4.5}
\end{multline}
assuming $\calD_{\humu(x,t)}$ and $\calD_{\hunu(x,t)}$to be nonspecial 
for $(x,t)\in\Omega$,  where $\Omega\subseteq\bbR^2$ is open and 
connected. We
refer to  Appendix \ref{A} for our notational conventions concerning Abel 
maps $\ua_{Q_{0}}$, $\ual_{Q_{0}}$ and $\theta$-functions.  Here 
$Q_{0}\in\calK_{n}\setminus\{\Pzeropm, \Pinfpm\}$ is a 
fixed base point which we will always choose among the branch points 
of $\calK_{n}$ (e.g., $Q_{0}=(E_{0},0)$). Indeed, by \eqref{T3.10}, 
\eqref{T4.3a}, \eqref{T4.4a}, and Theorem~\ref{taa17a}, $\phi(P,x,t)$ and 
\begin{equation}
\f{\theta\big(\uxi_{Q_{0}}-\ua_{Q_{0}}(P)
+\ual_{Q_{0}}(\calD_{\hunu(x,t)})\big)}
{\theta\big(\uxi_{Q_{0}}-\ua_{Q_{0}}(P)
+\ual_{Q_{0}}(\calD_{\humu(x,t)})\big)}\exp\left(\int_{Q_{0}}^P 
\omega_{\Pzerom,\Pinfm}^{(3)}\right) \lb{T4.5a}
\end{equation}
have the same singularity structure with respect to $P\in\calK_{n}$. 
Moreover, by \eqref{a37}, \eqref{a27}, and \eqref{aa51}, the expression 
\eqref{T4.5a} is single-valued and hence meromorphic on $\calK_{n}$. 
Nonspecialty of $\calD_{\humu(x,t)}$ and $\calD_{\hunu(x,t)}$ then 
yields \eqref{T4.5}.

It remains to
analyze the  function $C(x,t)$ in \eqref{T4.5} (which is $P$-independent) 
and in the course of that we will also obtain the theta function 
representations of $u$, $u^{*}$, $v$, $v^{*}$.  (The strategy to 
follow parallels the one used in \cite{GesztesyRatnaseelan:1996} in 
connection with algebro-geometric  solutions of the AKNS hierarchy.)

In the following it will occasionally be convenient to use a 
short-hand notation for the arguments of the theta functions in 
\eqref{T4.5} and hence we introduce the abbreviation
\begin{align}
\uz(P,\ul Q)=\uxi_{Q_{0}}-\ua_{Q_{0}}(P)+\ual_{Q_{0}}(\calD_{\ul Q}),\quad
\ul Q=(Q_{1}, \dots,Q_{n})\in\sigma^n \calK_{n}. \lb{T4.6}
\end{align}

Next we show that the Abel maps linearizes the auxiliary divisors 
$\calD_{\humu(x,t)}$ and $\calD_{\hunu(x,t)}$.

\begin{lemma}  \lb{lemmaT4.1}
Assume \eqref{T2.1}, \eqref{T2.8}, \eqref{T2.9}, and \eqref{T3.1}, 
and $(x,t), (x_{0},t_{0})\in\Omega$, where $\Omega\subseteq\bbR^2$ is 
open and connected.  Moreover, suppose $\calK_{n}$ is nonsingular 
and $\calD_{\humu(x,t)}$ and $\calD_{\hunu(x,t)}$ are  
nonspecial for $(x,t)\in\Omega$.  Then
\begin{align}
\ual_{Q_{0}}(\calD_{\humu(x,t)})&=\ual_{Q_{0}}(\calD_{\humu(x_{0},t_{0})})
+2i\ul c(n)(x-x_{0})-2i\ul c(1)g^{-1}_{n+1}(t-t_{0}), \lb{T4.7} \\
\ual_{Q_{0}}(\calD_{\hunu(x,t)})&=\ual_{Q_{0}}(\calD_{\hunu(x_{0},t_{0})})
+2i\ul c(n)(x-x_{0})-2i\ul c(1)g^{-1}_{n+1}(t-t_{0}). \lb{T4.8}
\end{align}
\end{lemma}
\begin{proof}  Given the expansions \eqref{b27} and 
\eqref{b27ab} of $\omega$ near $\Pinfpm$ 
and $\Pzeropm$, \eqref{T4.7} and \eqref{T4.8} are standard facts 
following from Lagrange interpolation results of the type (see, 
e.g., \cite{GesztesyHolden:1999c})
\begin{align}
\sum_{j=1}^n \f{\mu_{j}^{k-1}}{
\prod_{\substack{\ell=1\\ \ell\neq j}}^n(\mu_{j}-\mu_{\ell})}
&=\delta_{k,n}, \no \\
\sum_{j=1}^n \f{\mu_{j}^{k-1}\big(
\prod_{\substack{m=1\\ m\neq j}}^n \mu_{m} \big)}
{\prod_{\substack{\ell=1\\ \ell\neq  j}}^n(\mu_{j}-\mu_{\ell})}
&=(-1)^{n+1}\delta_{k,1}, \quad 
 k=1,\dots,n.\lb{T4.8a}     
\end{align}
\end{proof}

In Lemma \ref{lemmaT3.3} we determined the asymptotic behavior of 
$\phi(P,x,t)$ as $P\to \Pinfpm, \Pzeropm$ comparing 
\eqref{T3.3} with \eqref{T3.11} and \eqref{T3.12}.  Now we will 
recompute the asymtotics of $\phi$ starting from \eqref{T4.5}.

\begin{lemma}  \lb{lemmaT4.2}
Assume \eqref{T2.1}, \eqref{T2.8}, \eqref{T2.9}, and \eqref{T3.1}. 
Moreover,  suppose $\calK_{n}$ is nonsingular 
and $\calD_{\humu(x,t)}$ and $\calD_{\hunu(x,t)}$ are 
nonspecial for $(x,t)\in\Omega$, where $\Omega\subseteq\bbR^2$ is open 
and connected.  Then
\begin{align}
\begin{split}
\phi(P,x,t)&\underset{\zeta\to 0}{=} C(x,t) e^{\omega^{\infty_-}_{0}}
\f{\theta\big(\uz(\Pinfm,\hunu(x,t)) \big)}
{\theta\big(\uz(\Pinfm,\humu(x,t)) \big)}\, \zeta^{-1} \\
&\quad+C(x,t)\omega^\infty_{1}e^{\omega^{\infty_-}_{0}}
\f{\theta\big(\uz(\Pinfm,\hunu(x,t)) \big)}
{\theta\big(\uz(\Pinfm,\humu(x,t)) \big)} \\
&\quad-C(x,t)e^{\omega^{\infty_-}_{0}}\f{i}{2}\,\f{\partial}{\partial x}\left( 
\f{\theta\big(\uz(\Pinfm,\hunu(x,t)) \big)}
{\theta\big(\uz(\Pinfm,\humu(x,t)) \big)}\right)
+\Oh(\zeta) \text{  as $P\to \Pinfm$}, 
\end{split}\lb{T4.9} \\
\begin{split}
\phi(P,x,t)&\underset{\zeta\to 0}{=} C(x,t) e^{\omega^{\infty_+}_{0}}
\f{\theta\big(\uz(\Pinfp,\hunu(x,t)) \big)}
{\theta\big(\uz(\Pinfp,\humu(x,t)) \big)}\,  \\
&\quad-C(x,t)\omega^\infty_{1}e^{\omega^{\infty_+}_{0}}
\f{\theta\big(\uz(\Pinfp,\hunu(x,t)) \big)}
{\theta\big(\uz(\Pinfp,\humu(x,t)) \big)}\zeta \\
&\quad+C(x,t)e^{\omega^{\infty_+}_{0}}\f{i}{2}\,\f{\partial}{\partial x}\left( 
\f{\theta\big(\uz(\Pinfp,\hunu(x,t)) \big)}
{\theta\big(\uz(\Pinfp,\humu(x,t)) \big)}\right)\zeta
+\Oh(\zeta^2) \text{  as $P\to \Pinfp$}, 
\end{split}\lb{T4.10} \\
\begin{split}
\phi(P,x,t)&\underset{\zeta\to 0}{=} C(x,t) e^{\omega^{0,-}_{0}}
\f{\theta\big(\uz(\Pzerom,\hunu(x,t)) \big)}
{\theta\big(\uz(\Pzerom,\humu(x,t)) \big)}\, \zeta \\
&\quad+C(x,t)\omega^0_{1}e^{\omega^{0,-}_{0}}
\f{\theta\big(\uz(\Pzerom,\hunu(x,t)) \big)}
{\theta\big(\uz(\Pzerom,\humu(x,t)) \big)}\zeta^2 \\
&\quad+C(x,t)e^{\omega^{0,-}_{0}}\f{i}{2}\,\f{\partial}{\partial t}\left( 
\f{\theta\big(\uz(\Pzerom,\hunu(x,t)) \big)}
{\theta\big(\uz(\Pzerom,\humu(x,t)) \big)}\right)\zeta^2
+\Oh(\zeta^3) \text{  as $P\to \Pzerom$}, 
\end{split}\lb{T4.11} \\
\begin{split}
\phi(P,x,t)&\underset{\zeta\to 0}{=} C(x,t) e^{\omega^{0,+}_{0}}
\f{\theta\big(\uz(\Pzerop,\hunu(x,t)) \big)}
{\theta\big(\uz(\Pzerop,\humu(x,t)) \big)} \\
&\quad-C(x,t)\omega^0_{1}e^{\omega^{0,+}_{0}}
\f{\theta\big(\uz(\Pzerop,\hunu(x,t)) \big)}
{\theta\big(\uz(\Pzerop,\humu(x,t)) \big)}\zeta \\
&\quad-C(x,t)e^{\omega^{0,+}_{0}}\f{i}{2}\,\f{\partial}{\partial t}\left( 
\f{\theta\big(\uz(\Pzerop,\hunu(x,t)) \big)}
{\theta\big(\uz(\Pzerop,\humu(x,t)) \big)}\right)\zeta
+\Oh(\zeta^2) \text{  as $P\to \Pzerop$}. \end{split}\lb{T4.12} 
\end{align}
\end{lemma}
\begin{proof} Using \eqref{T4.13} and \eqref{T4.14} 
(cf.\ \eqref{b27} and \eqref{b27ab}) one obtains
\begin{align}
\uz(P,\humu(x,t))&\underset{\zeta\to 0}{=}\uz(\Pinfpm,\humu(x,t))
\mp \ul c(n) \zeta+\Oh(\zeta^2)
\text{  as $P\to \Pinfpm$}, \lb{T4.15}\\
\uz(P,\humu(x,t))&\underset{\zeta\to 0}{=}\uz(\Pzeropm,\humu(x,t))
\mp \ul c(1)g_{n+1}^{-1} \zeta+\Oh(\zeta^2)
\text{  as $P\to \Pzeropm$}\lb{T4.16}
\end{align}
and hence
\begin{align}
\begin{split}
\theta\big(\uz(P,\humu(x,t)) \big)&\underset{\zeta\to 0}{=}
\theta\big(\uz(\Pinfpm,\humu(x,t))\big) \\
&\qquad\pm
\f{i}{2}\, \f{\partial}{\partial x}
\theta\big(\uz(\Pinfpm,\humu(x,t))\big)\zeta+\Oh(\zeta^2) 
\text{  as $P\to \Pinfpm$}, 
\end{split}\lb{T4.17}\\
\begin{split}
\theta\big(\uz(P,\humu(x,t)) \big)&\underset{\zeta\to 0}{=}
\theta\big(\uz(\Pzeropm,\humu(x,t))\big)\\
&\qquad\mp\f{i}{2}\, \f{\partial}{\partial t}
\theta\big(\uz(\Pzeropm,\humu(x,t))\big)\zeta+\Oh(\zeta^2)
\text{  as $P\to \Pzeropm$}. \end{split}\lb{T4.18}
\end{align}
Here we used \eqref{T4.7} to convert the directional derivatives 
$\sum_{j=1}^n c_{j}(n)\partial/\partial w_{j}$ and 
$\sum_{j=1}^n c_{j}(1)\partial/\partial w_{j}$, $\ul 
w=(w_{1},\dots,w_{n})\in\bbC^n$ into $\partial/\partial x$ and 
$\partial/\partial t$ derivatives.  Since by \eqref{T4.8} exactly the 
same formulas \eqref{T4.17} and \eqref{T4.18} apply to  
$\calD_{\hunu(x,t)}$, insertion of 
\eqref{T4.3a}, \eqref{T4.4a},  \eqref{T4.17},  and 
\eqref{T4.18} (and their $\calD_{\hunu(x,t)}$ analogs) into 
\eqref{T4.5} proves \eqref{T4.9}--\eqref{T4.12}. 
\end{proof}

Lemma \ref{lemmaT4.2} may seem to be just another asymptotic result, 
however, a comparison with Lemma \ref{lemmaT3.3} reveals that in 
passing we have actually derived the theta function representations 
for $u$, $v$, $u^{*}$, and $v^*$.

\begin{theorem}\lb{theoremT4.3}
Assume \eqref{T2.1}, \eqref{T2.8}, \eqref{T2.9}, and \eqref{T3.1}, 
and suppose $\calK_{n}$ is nonsingular.  In addition, let 
$P\in\calK_{n}$ and $(x,t)\in\Omega$, where $\Omega\subseteq\bbR^2$ is 
open and connected.  Moreover, suppose that $\calD_{\humu(x,t)}$ 
and $\calD_{\hunu(x,t)}$ are nonspecial for 
$(x,t)\in\Omega$.  Then $\phi(P,x,t)$ admits the representation
\begin{equation}
\phi(P,x,t)=C_{0}e^{2i(\omega_{1}^\infty x-\omega_{1}^0 t)} \,
\f{\theta\big(\uz(P,\hunu(x,t)) \big)}{\theta\big(\uz(P,\humu(x,t)) \big)}
\exp\left(\int_{Q_{0}}^{P} \omega_{\Pzerom,\Pinfm}^{(3)} \right)
\lb{T4.19}
\end{equation}
for some constant $C_{0}\in\bbC\setminus\{0\}$ and the theta function 
representations for the algebro-geometric solutions $u$, $u^{*}$, $v$, 
and $v^{*}$ of the classical massive Thirring system 
\eqref{T2.22}--\eqref{T2.25} read
\begin{align}
u(x,t)&=-C_{0}^{-1} e^{-\omega_{0}^{0,+}}
\f{\theta\big(\uz(\Pzerop,\humu(x,t)) \big)}
{\theta\big(\uz(\Pzerop,\hunu(x,t)) \big)}
e^{-2i(\omega_{1}^\infty x-\omega_{1}^0 t)}, \lb{T4.20} \\
v(x,t)&=-C_{0}^{-1} e^{-\omega_{0}^{\infty_-}}
\f{\theta\big(\uz(\Pinfm,\humu(x,t)) \big)}
{\theta\big(\uz(\Pinfm,\hunu(x,t)) \big)}
e^{-2i(\omega_{1}^\infty x-\omega_{1}^0 t)}, \lb{T4.22} \\
u^{*}(x,t)&=C_{0} e^{\omega_{0}^{0,-}}
\f{\theta\big(\uz(\Pzerom,\hunu(x,t)) \big)}
{\theta\big(\uz(\Pzerom,\humu(x,t)) \big)}
e^{2i(\omega_{1}^\infty x-\omega_{1}^0 t)}, \lb{T4.21} \\
v^{*}(x,t)&=C_{0} e^{\omega_{0}^{\infty_+}}
\f{\theta\big(\uz(\Pinfp,\hunu(x,t)) \big)}
{\theta\big(\uz(\Pinfp,\humu(x,t)) \big)}
e^{2i(\omega_{1}^\infty x-\omega_{1}^0 t)}, \lb{T4.23} 
\end{align}
with $\omega_0^{0,\pm}, \omega_1^0, \omega_0^{\infty_\pm}$, and 
$\omega_1^\infty$ given by \eqref{T4g}--\eqref{T4n} {\rm (}cf. also 
\eqref{T4.3a}--\eqref{T4.4c}{\rm )}.
\end{theorem}
\begin{proof}
A comparison of \eqref{T3.27}--\eqref{T3.30} and 
\eqref{T4.9}--\eqref{T4.12} yields
\begin{equation}
    C_{x}(x,t)=2i\omega_{1}^\infty C(x,t), \quad
    C_{t}(x,t)=-2i\omega_{1}^0 C(x,t), \lb{T4.24}
\end{equation}
and hence
\begin{equation}
    C(x,t)=C_{0}e^{2i(\omega_{1}^\infty x-\omega_{1}^0 t)}, \lb{T4.25}
\end{equation}
proves \eqref{T4.19}.  Insertion of \eqref{T4.25} into the leading 
asymptotic term of \eqref{T4.9}--\eqref{T4.12} then yields 
\eqref{T4.20}--\eqref{T4.23}.
\end{proof}

\begin{remark} \lb{remarkT4.4}
(i) The constant $C_{0}$ in \eqref{T4.19}--\eqref{T4.23} remains open 
due to the scaling invariance \eqref{T2.27} of the Thirring system. 
One can rewrite \eqref{T4.20}--\eqref{T4.23} in the form
\begin{align}
\begin{split}
u(x,t)&= u(x_{0},t_{0})
\f{\theta\big(\uz(\Pzerop,\humu(x,t)) \big)
\theta\big(\uz(\Pzerop,\hunu(x_{0},t_{0})) \big)}
{\theta\big(\uz(\Pzerop,\humu(x_{0},t_{0})) \big)
\theta\big(\uz(\Pzerop,\hunu(x,t)) \big)}\times \\
&\quad
\times\exp\left(-2i(\omega_{1}^\infty(x-x_{0})-\omega_{1}^0(t-t_{0}))  
\right), 
\end{split}\lb{T4.26} \\
\begin{split}
v(x,t)&= v(x_{0},t_{0})
\f{\theta\big(\uz(\Pinfm,\humu(x,t)) \big)
\theta\big(\uz(\Pinfm,\hunu(x_{0},t_{0})) \big)}
{\theta\big(\uz(\Pinfm,\humu(x_{0},t_{0})) \big)
\theta\big(\uz(\Pinfm,\hunu(x,t)) \big)}\times \\
&\quad
\times\exp\left(-2i(\omega_{1}^\infty(x-x_{0})-\omega_{1}^0(t-t_{0}))  
\right), \end{split}\lb{T4.28} \\
\begin{split}
u^{*}(x,t)&= u^{*}(x_{0},t_{0})
\f{\theta\big(\uz(\Pzerom,\hunu(x,t)) \big)
\theta\big(\uz(\Pzerom,\humu(x_{0},t_{0})) \big)}
{\theta\big(\uz(\Pzerom,\hunu(x_{0},t_{0})) \big)
\theta\big(\uz(\Pzerom,\humu(x,t)) \big)}\times \\
&\quad \times\exp\left(2i(\omega_{1}^\infty(x-x_{0})-\omega_{1}^0(t-t_{0}))  
\right), 
\end{split}\lb{T4.27} \\
\begin{split}
v^{*}(x,t)&= v^{*}(x_{0},t_{0})
\f{\theta\big(\uz(\Pinfp,\hunu(x,t)) \big)
\theta\big(\uz(\Pinfp,\humu(x_{0},t_{0})) \big)}
{\theta\big(\uz(\Pinfp,\hunu(x_{0},t_{0})) \big)
\theta\big(\uz(\Pinfp,\humu(x,t)) \big)}\times \\
&\quad \times\exp\left(2i(\omega_{1}^\infty(x-x_{0})-\omega_{1}^0(t-t_{0}))  
\right), 
\end{split}\lb{T4.29} 
\end{align}
where
\begin{align}
\uz(Q,\humu(x,t))&=\uz(Q,\humu(x_{0},t_{0}))+2i\ul c(n) (x-x_{0})
+2i\ul c(1)g_{n+1}^{-1}(t-t_{0}),\lb{T4.30} \\
\uz(Q,\hunu(x,t))&=\uz(Q,\hunu(x_{0},t_{0}))+2i\ul c(n) (x-x_{0})
+2i\ul c(1)g_{n+1}^{-1}(t-t_{0}),\lb{T4.31} 
\end{align}
by \eqref{T4.6}, \eqref{T4.7}, and \eqref{T4.8}. \\
(ii) Since the divisors $\calD_{\Pzerom\hunu(x,t)}$ and 
$\calD_{\Pinfm\humu(x,t)}$ are linearly independent by 
\eqref{T3.10}, one infers
\begin{align}
\ual_{Q_{0}}(\calD_{\hunu(x,t)})=\ual_{Q_{0}}(\calD_{\humu(x,t)})+
\ul\Delta, \quad (x,t)\in\Omega, \quad 
\ul\Delta=\ua_{\Pzerom}(\Pinfm).  \lb{T4.32}    
\end{align}
Hence on can replace $\uz(Q,\hunu(x,t))$ in 
\eqref{T4.19}--\eqref{T4.23}, \eqref{T4.26}--\eqref{T4.29}, 
\eqref{T4.31} in terms of $\uz(Q,\humu(x,t))$ according to 
\begin{equation}
\uz(Q,\hunu(x,t))=\uz(Q,\humu(x,t))+\ul\Delta. \lb{T4.33}    
\end{equation}
\end{remark}

In principle, Theorem \ref{theoremT4.3} completes the primary aim of 
this paper, the derivation of the theta function representation of
algebro-geometric  solutions of the classical massive Thirring system 
\eqref{T2.22}--\eqref{T2.25}.  The reader will have noticed that our 
approach thus far is nontraditional in the sense that we did not use 
the Baker--Akhiezer vector $\Psi$ at all, but instead put all emphasis 
on the meromorphic $\phi$ on $\calK_{n}$.  Just for completeness we 
finally derive the theta function representation for $\psi_{1}$ in 
\eqref{T3.17}.

The singularity structure of $\psi_{1}(P,x,x_{0},t,t_{0})$
near $\Pinfpm$ displayed in  Lemma \ref{lemmaT3.4} suggests
introducing Abelian differentials 
$\omega^{(2)}_{Q,0}$ of the second kind, normalized by the vanishing 
of their $a$-periods,
\begin{equation}
\int_{a_{j}}\omega^{(2)}_{Q,0}=0, \quad j=1,\dots,n, 
\lb{T4.34}    
\end{equation}
with a second-order pole at $Q$ of the type 
\begin{equation}
\omega^{(2)}_{Q,0}\underset{\zeta\to 0}{=}(\zeta^{-2}+\Oh(1))d\zeta
\text{  as $P\to Q$}, 
\lb{T4.35}    
\end{equation}
and holomorphic on $\calK_{n}\setminus\{Q\}$.  More precisely, we 
introduce
\begin{align}
\Omega^{(2)}_{\infty,0}&=\omega^{(2)}_{\Pinfp,0}-\omega^{(2)}_{\Pinfm,0}, 
\lb{T4.36} \\
\Omega^{(2)}_{0,0}&=\omega^{(2)}_{\Pzerop,0}-\omega^{(2)}_{\Pzerom,0}, 
\lb{T4.37}
\end{align}
and note that
\begin{align}
\int_{Q_{0}}^{P}\Omega^{(2)}_{\infty,0}&\underset{\zeta\to 0}{=}
\pm(\zeta^{-1}+e_{\infty,0}+e_{\infty,1}\zeta+\Oh(\zeta^2))
\text{  as $P\to\Pinfmp$},\lb{T4.38}\\
\int_{Q_{0}}^{P}\Omega^{(2)}_{0,0}&\underset{\zeta\to 0}{=}
\pm(\zeta^{-1}+e_{0,0}+e_{0,1}\zeta+\Oh(\zeta^2))
\text{  as $P\to\Pzeromp$}.\lb{T4.39}
\end{align}

\begin{theorem} \lb{theoremT4.5}
Assume \eqref{T2.1}, \eqref{T2.8}, \eqref{T2.9}, \eqref{T3.1}, 
and suppose $\calK_{n}$ is nonsingular.  In addition, let 
$P\in\calK_{n}\setminus\{\Pzeropm,\Pinfpm\}$ and $(x,t), 
(x_{0},t_{0})\in\Omega$, where $\Omega\subseteq\bbR^2$ is open and 
connected.  Moreover, suppose that $\calD_{\humu(x,t)}$ and 
$\calD_{\hunu(x,t)}$ are nonspecial 
for $(x,t)\in\Omega$.  Then $\psi_{1}(P,x,x_{0},t,t_{0})$ admits the 
representation
\begin{multline}
\psi_{1}(P,x,x_{0},t,t_{0})=\left(
\f{\theta\big(\uz(\Pinfm,\humu(x_{0},t_{0}))  \big)
   \theta\big(\uz(\Pinfp,\hunu(x_{0},t_{0}))  \big) }
   {\theta\big(\uz(\Pinfp,\humu(x,t))  \big)
   \theta\big(\uz(\Pinfm,\hunu(x,t))  \big)}
\right)^{1/2} \, \times \\
\times\f{\theta\big(\uz(P,\humu(x,t))\big)}
{\theta\big(\uz(P,\humu(x_{0},t_{0}))\big)} \times \\
\times \exp\left(-i(x-x_{0})\bigg(\omega_{1}^\infty+\int_{Q_{0}}^P 
\Omega^{(2)}_{\infty,0}\bigg) 
  +i(t-t_{0})\bigg(\omega_{1}^0
+\int_{Q_{0}}^P \Omega^{(2)}_{0,0}\bigg)\right), 
   \lb{T4.41}
\end{multline}
or, equivalently,
\begin{multline}
\psi_{1}(P,x,x_{0},t,t_{0})=\left(
\f{\theta\big(\uz(\Pzerom,\humu(x_{0},t_{0}))  \big)
   \theta\big(\uz(\Pzerop,\hunu(x_{0},t_{0}))  \big) }
   {\theta\big(\uz(\Pzerom,\humu(x,t))  \big)
   \theta\big(\uz(\Pzerop,\hunu(x,t))  \big)}
\right)^{1/2} \, \times \\
 \times\f{\theta\big(\uz(P,\humu(x,t))\big)}
{\theta\big(\uz(P,\humu(x_{0},t_{0}))\big)} \times \\
\times \exp\left(-i(x-x_{0})\bigg(\omega_{1}^\infty+\int_{Q_{0}}^P 
\Omega^{(2)}_{\infty,0}\bigg) 
 +i(t-t_{0})\bigg(\omega_{1}^0
+\int_{Q_{0}}^P \Omega^{(2)}_{0,0}\bigg)\right). 
   \lb{T4.41A}
\end{multline}
\end{theorem}
\begin{proof} 
Introducing
\begin{multline}
\hat \psi_{1}(P,x,x_{0},t,t_{0})=\f{C(x,t)}{C(x_{0},t_{0})} \,
\f{\theta\big(\uz(P,\humu(x,t))\big)}
{\theta\big(\uz(P,\humu(x_{0},t_{0}))\big)}\times\\
\times\exp\left(-i(x-x_{0})\int_{Q_{0}}^P 
\Omega^{(2)}_{\infty,0}+i(t-t_{0})\int_{Q_{0}}^P
\Omega^{(2)}_{0,0}\right), \\
 P\in\calK_{n}\setminus\{\Pzeropm,\Pinfpm\}, \, (x,t), 
(x_{0},t_{0})\in\Omega,
\lb{T4.40}
\end{multline}
with an appropriate normalization $C(x,t)$ (which is $P$-independent)
to be determined later, we next intend to prove that
\begin{equation}
\psi_{1}(P,x,x_{0},t,t_{0})=\hat \psi_{1}(P,x,x_{0},t,t_{0}), \quad 
P\in\calK_{n}\setminus\{\Pzeropm,\Pinfpm\}, \, (x,t), 
(x_{0},t_{0})\in\Omega. \lb{T4.40a}
\end{equation}
A comparison of \eqref{T3.21}, \eqref{T3.34a}, \eqref{T3.34b},
\eqref{T4.38}, \eqref{T4.39}, and \eqref{T4.40} shows that $\psi_1$
and $\hat\psi_1$ share the identical essential singularity near
$\Pinfpm$. Next we turn to the local bahavior of
$\psi_{1}(P,x,x_{0},t,t_{0})$ with respect to its zeros and poles.
We temporarily restrict $\Omega$ to $\ti\Omega\subseteq\Omega$ such
that for all $(x^{\prime},s)\in\ti\Omega$,
$\mu_j(x^{\prime},s)\neq\mu_k(x^{\prime},s)$ for all $j\neq k$,
$j,k=1,\dots,n$. Then arguing as in the paragraph following
\eqref{T3.26} one infers from \eqref{T3.17} that 
\begin{align} 
&\psi_1(P,x,x_{0},t,t_{0})=\begin{cases}
(\mu_j(x,t)-z)\Oh(1) & \text{ as }
P\to\hmu_j(x,t)\neq\hmu(x_{0},t_{0}),  \\
\Oh(1) & \text{ as } P\to\hmu_j(x,t)=\hmu(x_{0},t_{0}),  \\
(\mu_j(x_{0},t_{0})-z)^{-1}\Oh(1) & \text{ as }
P\to\hmu_j(x_{0},t_{0})\neq\hmu(x,t),  \end{cases} \no \\
& \hspace*{5cm} P=(z,y)\in\calK_{n}, \,\, (x,t),
(x_{0},t_{0})\in\ti\Omega, \lb{T4.40b}
\end{align}
where $\Oh(1)\neq 0$. Applying Lemma~\ref{la6} then proves
\eqref{T4.40a} for $(x,t), (x_{0},t_{0})\in\ti\Omega$. By continuity
this extends to 
$(x,t), (x_{0},t_{0})\in\Omega$ as long as
$\calD_{\humu(x,t)}\in\sigma^{n}\calK_{n}$ remains nonspecial.
Finally we determine $C(x,t)/C(x_{0},t_{0})$. A comparison of
\eqref{T2.5}, \eqref{T2.21}, 
\eqref{T3.22}, \eqref{T4.26}, \eqref{T4.28}, and \eqref{T4.40} yields
\begin{align}
&\psi_{1}(P,x,x_{0},t,t_{0})\psi_{1}(P^{*},x,x_{0},t,t_{0}) \no \\
&\quad\underset{z\to\infty}{=} \f{C(x,t)^2}{C(x_{0},t_{0})^2}
\f{\theta\big(\uz(\Pinfp,\humu(x,t))  \big)
   \theta\big(\uz(\Pinfm,\humu(x,t))  \big) }
   {\theta\big(\uz(\Pinfp,\humu(x_{0},t_{0}))  \big)
   \theta\big(\uz(\Pinfm,\humu(x_{0},t_{0}))  \big)}\times \no \\
&\quad\qquad\times\exp\left(-2i((x-x_{0})e_{\infty,0}
-(t-t_{0}) e_{0,0}) \right) \no \\
&\quad=\f{v(x,t)}{v(x_{0},t_{0})} \lb{T4.42} \\
&\quad=\f{\theta\big(\uz(\Pinfm,\humu(x,t))  \big)
   \theta\big(\uz(\Pinfm,\hunu(x_{0},t_{0}))  \big) }
   {\theta\big(\uz(\Pinfm,\humu(x_{0},t_{0}))  \big)
   \theta\big(\uz(\Pinfm,\hunu(x,t))  \big)}
   \exp\left(-2i((x-x_{0})\omega_{1}^\infty-(t-t_{0})\omega_{1}^0) 
\right)
\no 
\end{align}
and
\begin{align}
&\psi_{1}(P,x,x_{0},t,t_{0})\psi_{1}(P^{*},x,x_{0},t,t_{0}) \no \\
&\qquad\underset{z\to 0}{=} \f{C(x,t)^2}{C(x_{0},t_{0})^2}
\f{\theta\big(\uz(\Pzerop,\humu(x,t))  \big)
   \theta\big(\uz(\Pzerom,\humu(x,t))  \big) }
   {\theta\big(\uz(\Pzerop,\humu(x_{0},t_{0}))  \big)
   \theta\big(\uz(\Pzerom,\humu(x_{0},t_{0}))  \big)}\times \no \\
&\qquad\qquad\times 
\exp\left(-2i((x-x_{0})e_{\infty,0}-(t-t_{0}) e_{0,0}) \right) \no \\
&\qquad=\f{u(x,t)}{u(x_{0},t_{0})} \lb{T4.43} \\
&\qquad=\f{\theta\big(\uz(\Pzerop,\humu(x,t))  \big)
   \theta\big(\uz(\Pzerop,\hunu(x_{0},t_{0}))  \big) }
   {\theta\big(\uz(\Pzerop,\humu(x_{0},t_{0}))  \big)
   \theta\big(\uz(\Pzerop,\hunu(x,t))  \big)}\times \no \\
&\qquad\qquad\times  
\exp\left(-2i((x-x_{0})\omega_{1}^\infty-(t-t_{0})\omega_{1}^0) 
\right).  \no   
\end{align}
Thus, \eqref{T4.42} implies
\begin{multline}
\f{C(x,t)^2}{C(x_{0},t_{0})^2}=
\f{\theta\big(\uz(\Pinfp,\humu(x_{0},t_{0}))  \big)
   \theta\big(\uz(\Pinfm,\hunu(x_{0},t_{0}))  \big) }
   {\theta\big(\uz(\Pinfp,\humu(x,t))  \big)
   \theta\big(\uz(\Pinfm,\hunu(x,t))  \big)}\times \\
    \times\exp\left(-2i((x-x_{0})\omega_{1}^\infty 
   -(t-t_{0})\omega^0_{1}) \right) \lb{T4.44}
\end{multline}
and \eqref{T4.43} yields
\begin{multline}
\f{C(x,t)^2}{C(x_{0},t_{0})^2}=
\f{\theta\big(\uz(\Pzerom,\humu(x_{0},t_{0}))  \big)
   \theta\big(\uz(\Pzerop,\hunu(x_{0},t_{0}))  \big) }
   {\theta\big(\uz(\Pzerom,\humu(x,t))  \big)
   \theta\big(\uz(\Pzerop,\hunu(x,t))  \big)}\times \\
   \times\exp\left(-2i((x-x_{0})\omega_{1}^\infty 
   -(t-t_{0})\omega^0_{1}) \right). \lb{T4.45}
\end{multline}
In order to reconcile the two expressions \eqref{T4.44} and 
\eqref{T4.45} for $C(x,t)^2/C(x_{0},t_{0})^2$ it suffices to recall 
the linear dependence of the divisors $\calD_{\Pinfm \humu(x,t)}$ and 
$\calD_{\Pzerom \hunu(x,t)}$, that is,
\begin{align}
    \ua_{Q_{0}}(\Pinfm)&+\ual_{Q_{0}}(\calD_{\humu(x,t)})
    =\ua_{Q_{0}}(\Pzerom)+\ual_{Q_{0}}(\calD_{\hunu(x,t)}), \lb{T4.46}\\
\intertext{and}
\ua_{Q_{0}}(\Pzerom)&=-\ua_{Q_{0}}(\Pzerop), \quad 
\ua_{Q_{0}}(\Pinfm)=-\ua_{Q_{0}}(\Pinfp), \lb{T4.47}
\end{align}
to conclude that
\begin{equation}
    \uz(\Pinfp,\humu(x,t))= \uz(\Pzerop,\hunu(x,t)), \quad
    \uz(\Pzerom,\humu(x,t))= \uz(\Pinfm,\hunu(x,t)) \lb{T4.48}
\end{equation}
and hence equality of the right-hand sides of \eqref{T4.44} and 
\eqref{T4.45}.  This proves \eqref{T4.41} and \eqref{T4.41A}.
\end{proof}

The explicit representation \eqref{T4.41} for $\psi_{1}$ complements 
Lemma \ref{lemmaT3.2} and shows that $\psi_{1}$ stays meromorphic 
on $\calK_{n}\setminus\{\Pinfpm\}$ as long as $\calD_{\humu(x,t)}$ and
$\calD_{\hunu(x,t)}$  are nonspecial (assuming $\calK_{n}$ 
to be  nonsingular).  An analogous theta function derivation can be
performed  for $\zeta\psi_{2}(P,\zeta,x,x_{0},t,t_{0})$, but we omit 
further details at this point. 

We emphasize that $\phi(P,x,t)$ and 
$\psi_{1}(P,x,x_{0},t,t_{0})$ are naturally defined on the two-sheeted 
Riemann surface $\calK_{n}$, whereas 
$\psi_{2}(P,\zeta,x,x_{0},t,t_{0})$ requires a four-sheeted Riemann 
surface due to the additional factor $1/z^{1/2}$ in \eqref{T3.18}.  In 
particular, the Baker--Akhiezer vector $\Psi(P,\zeta,x,x_{0},t,t_{0})$ 
in \eqref{T3.16} requires a four-sheeted Riemann surface, clearly a 
disadvantage when compared to our use of $\phi(P,x,t)$.  Finally, we 
note that reality constraints of  the type \eqref{T2.26} and their 
effects on algebro-geometric quantities, such as pairs of real and 
complex conjugate branch points of $\calK_{n}$, etc., are discussed 
in \cite{Bikbaev:1985} (see also \cite{Date:1982}).

We conclude with the elementary genus zero example (i.e., $n=0$), a
case thus far excluded in this section.

\begin{example}\lb{exampleT4.7}
Assume $n=0$. Then 
\begin{align}
\calK_{0} \colon \calF_0(z,y)&=y^2-R_2(z)=y^2-(z-E_0)(z-E_1)=0, \\
c_1&=-(E_0+E_1)/2, \quad g_1=(E_0E_1)^{1/2},\\
\omega^\infty_1&=(g_1+c_1)/2, \quad
\omega^0_1=-\omega^\infty_1/(E_0E_1), \\
v(x,t)&=v(x_0,t_0)\exp(-2i(\omega^\infty_1(x-x_0)
-\omega^0_1(t-t_0))) \no \\
&=g_1 u(x,t),\\
v^{*}(x,t)&=v^{*}(x_0,t_0)\exp(2i(\omega^\infty_1(x-x_0)
-\omega^0_1(t-t_0)))\no \\
&=g_1 u^{*}(x,t),  \\
v(x,t)v^{*}(x,t)&=(c_1-g_1)/2=g_1^2 u(x,t)u^{*}(x,t), \\
\phi(P,x,t)&=\f{y(P)+z+g_1}{-2v(x,t)}
=\f{-2v^{*}(x,t)z}{y(P)-z-g_1}, \\
\psi_1(P,x,x_0,t,t_0)&=\exp(-i(x-x_0)(y(P)+\omega^\infty_1)
-i(t-t_0)(g_1^{-1}z^{-1}y(P)-\omega^0_1)).  
\end{align} 
\end{example}

\appendix
\section{Hyperelliptic curves and their theta functions}\lb{A}
\renewcommand{\theequation}{A.\arabic{equation}}
\renewcommand{\thetheorem}{A.\arabic{theorem}}
\setcounter{theorem}{0}
\setcounter{equation}{0}

We give a brief summary of some of the fundamental properties
and notations needed from the
theory of hyperelliptic curves.  More details can be found in
some of the standard textbooks
\cite{FarkasKra:1992} and \cite{Mumford:1984}, as well as monographs
dedicated to integrable systems such as 
\cite{BelokolosBobenkoEnolskiiItsMatveev:1994}, Ch.\ 2,
\cite{GesztesyHolden:2000}, App. A--C.

Fix $\N\in\bbN$. The hyperelliptic curve $\calK_\N$
of genus $\N$ used in Sections~\ref{Ts3} and \ref{Ts4} is 
defined by
\begin{multline}
\calK_\N: \, \calF_\N(z,y)=y^2-R_{2\N+2}(z)=0, \quad
R_{2\N+2}(z)=\prod_{m=0}^{2\N+1}(z-E_m), \\
\quad \{E_m\}_{m=0,\dots,2\N+1}\subset\bbC, \quad
E_m \neq E_{n} \text{ for } m \neq n, \, m,n=0,\dots,2\N+1.
\label{b1}
\end{multline}
The curve \eqref{b1} is compactified by adding the
points $P_{\infty_+}$ and $P_{\infty_-}$, 
$P_{\infty_+} \neq P_{\infty_-}$, at infinity.
One then introduces an appropriate set of
$\N+1$ nonintersecting cuts $\calC_j$ joining
$E_{m(j)}$ and $E_{m^\prime(j)}$. We denote
\begin{equation}
\calC=\bigcup_{j\in \{1,\dots,\N+1 \}}\calC_j,
\quad
\calC_j\cap\calC_k=\emptyset,
\quad j\neq k.\label{b2}
\end{equation}
Define the cut plane
\begin{equation}
\Pi=\bbC\setminus\calC, \label{b3}
\end{equation}
and introduce the holomorphic function
\begin{equation}
R_{2\N+2}(\dott)^{1/2}\colon \Pi\to\bbC, \quad
z\mapsto \left(\prod_{m=0}^{2\N+1}(z-E_m) \right)^{1/2}\label{b4}
\end{equation}
on $\Pi$ with an appropriate choice of the square root
branch in \eqref{b4}. Define
\begin{equation}
\calM_{\N}=\{(z,\sigma R_{2\N+2}(z)^{1/2}) \mid 
z\in\bbC,\; \sigma\in\{\pm 1\}
\}\cup\{P_{\infty_+},P_{\infty_-}\} \label{b5}
\end{equation}
by extending $R_{2\N+2}(\dott)^{1/2}$ to $\calC$. The
hyperelliptic curve $\calK_\N$ is then the set
$\calM_{\N}$ with its natural complex structure obtained
upon gluing the two sheets of $\calM_{\N}$
crosswise along the cuts. The set of branch points 
$\calB(\calK_\N)$ of $\calK_\N$ is given by
\begin{equation}
\calB(\calK_\N)=\{(E_m,0)\}_{m=0,\dots,2\N+1} \lb{5a}
\end{equation}
and finite points $P$ on $\calK_\N$ are denoted by
$P=(z,y)$, where $y(P)$ denotes the meromorphic function
on $\calK_\N$ satisfying $\calF_\N(z,y)=y^2-R_{2\N+2}(z)=0$. 
Local coordinates near $P_0=(z_0,y_0)\in\calK_\N\setminus
\{\calB(\calK_\N)\cup\{P_{\infty_+},P_{\infty_-}\}\}$ are 
given by $\zeta_{P_0}=z-z_0$, near $P_{\infty_\pm}$ by 
$\zeta_{P_{\infty_\pm}}=1/z$, and near branch points
$(E_{m_0},0)\in\calB(\calK_\N)$ by 
$\zeta_{(E_{m_0},0)}=(z-E_{m_0})^{1/2}$. The Riemann surface  
$\calK_\N$ defined in this manner has topological genus $\N$.

One verifies that $dz/y$ is a holomorphic differential
on $\calK_\N$ with zeros of order $\N-1$ at $P_{\infty_\pm}$ 
and hence
\begin{equation}
\eta_j=\frac{z^{j-1}dz}{y}, \quad j=1,\dots,\N
\lb{b24}
\end{equation}
form a basis for the space of holomorphic differentials
on $\calK_\N$.  Introducing the
invertible matrix $C$ in $\bbC^\N$,
\begin{align}
\begin{split}
C & =(C_{j,k})_{j,k=1,\dots,\N}, \quad C_{j,k}
= \int_{a_k} \eta_j, \\
\underline{c} (k) & = (c_1(k), \dots,
c_n(k)), \quad c_j (k) =
C_{j,k}^{-1},
\lb{A.7}
\end{split}
\end{align}
the corresponding basis of normalized holomorphic 
differentials $\omega_j$, $j=1,\dots,\N$ on $\calK_\N$ is given by 
\begin{equation}
\omega_j = \sum_{\ell=1}^\N c_j (\ell) \eta_\ell,
\quad \int_{a_k} \omega_j =
\delta_{j,k}, \quad j,k=1,\dots,\N. \lb{b26}
\end{equation}
Here $\{a_j,b_j\}_{j=1,\dots,\N}$ is a homology basis for 
$\calK_\N$ with intersection matrix of the cycles satisfying
\begin{equation}
a_j \circ b_k=\delta_{j,k}, \quad j,k=1,\dots,\N. \lb{c26}
\end{equation}

Near $P_{\infty_\pm}$ one infers
\begin{align}
\ul\omega & = (\omega_1,\dots,\omega_\N)=
\pm \bigg( \sum_{j=1}^\N \f{\ul c (j)
\zeta^{\N-j}}{\big(\prod_{m=0}^{2\N+1}
(1-E_m \zeta) \big)^{1/2}} \bigg) d\zeta \no \\
& \underset{\zeta \to 0}{=} \pm \bigg( \ul c (\N) +
\bigg( \frac12 \ul c (\N)
\sum_{m=0}^{2\N+1} E_m +\ul c
(\N-1) \bigg) \zeta + \Oh(\zeta^2) \bigg)d\zeta, 
\quad \zeta=1/z, \lb{b27} 
\end{align}
and 
\begin{equation}
y(P) \underset{\zeta \to 0}{=} \mp \bigg(1-\frac12
\bigg( \sum_{m=0}^{2\N+1} E_m \bigg)\zeta +
\Oh(\zeta^2)\bigg)\zeta^{-\N-1} \text{ as }
P\to P_{\infty_\pm}. \lb{b27a}
\end{equation}
Similarly, near $\Pzeropm$ one computes
\begin{align}
&\ul\omega \underset{\zeta\to 0}{=} \pm \f{1}{g_{n+1}}\bigg(\ul c(1)
+\bigg(\f{1}{2}\ul c(1)\sum_{m=0}^{2n+1}E_m^{-1}+\ul c(2)\bigg)\zeta 
+ \Oh(\zeta^2) \bigg)d\zeta, \lb{b27ab} \\
& \hspace*{4.5cm} g_{n+1}=\bigg(\prod_{m=0}^{2\N+1}E_m\bigg)^{1/2},
\quad \zeta=z,  \no
\end{align}
using
\begin{equation}
y(P)\underset{\zeta\to 0}{=} \pm g_{n+1} + \Oh(\zeta)
\text{ as } P\to\Pzeropm, \lb{b27ac}
\end{equation}
with the sign of $g_{n+1}$ determined by the compatibility 
of charts. 

Associated with the homology basis
$\{a_j, b_j\}_{j=1,\dots,\N}$ we
also recall the canonical dissection of $\calK_\N$
along its cycles yielding
the simply connected interior $\hatt \calK_\N$ of the
fundamental polygon $\partial {\hatt \calK}_\N$ given by
\begin{equation}
\partial  {\hatt \calK}_\N =a_1 b_1 a_1^{-1} b_1^{-1}
a_2 b_2 a_2^{-1} b_2^{-1} \cdots
a_\N^{-1} b_\N^{-1}.
\lb{a25}
\end{equation}
Let $\calM (\calK_\N)$ and $\calM^1 (\calK_\N)$ denote the
set of meromorphic
functions (0-forms) and meromorphic
differentials (1-forms)
on $\calK_\N$. The residue of a meromorphic differential
$\nu\in \calM^1 (\calK_\N)$ at a
point $Q \in \calK_\N$ is defined by
\begin{equation}
\text{res}_{Q}(\nu)
=\frac{1}{2\pi i} \int_{\gamma_{Q}} \nu,
\lb{a33}
\end{equation}
where $\gamma_{Q}$ is a counterclockwise oriented
smooth simple closed
contour encircling $Q$ but no other pole of
$\nu$.  Holomorphic
differentials are also called Abelian differentials
of the first kind (dfk). Abelian differentials of the 
second kind
(dsk) $\omega^{(2)} \in \calM^1 (\calK_\N)$ are characterized
by the property
that all their residues vanish.  They are
normalized, for
instance, by demanding that all their $a$-periods
vanish, that is,
\begin{equation}
\int_{a_j} \omega^{(2)} =0, \quad  j=1,\dots,\N.
\lb{a34}
\end{equation}
If $\omega_{P_1, n}^{(2)}$ is a dsk on $\calK_\N$ whose
only pole is $P_1 \in \hatt \calK_\N$ with principal part
$\zeta^{-n-2}\,d\zeta$, $n\in\bbN_0$ near
$P_1$ and $\omega_j =
 (\sum_{m=0}^\infty d_{j,m} (P_1) \zeta^m)\, d\zeta$
near $P_1$, then
\begin{equation}
\frac{1}{2\pi i} \int_{b_j} \omega_{P_1, m}^{(2)} =
 \frac{d_{j,m} (P_1)}{m+1}, \quad m=0,1,\dots
\lb{a35}
\end{equation}

Any meromorphic differential $\omega^{(3)}$ on
$\calK_\N$ not of the first or
second kind is said to be of the third
kind (dtk).
A dtk $\omega^{(3)} \in \calM^1 (\calK_\N)$
is usually normalized by the vanishing of its
$a$-periods, that is,
\begin{equation}
\int_{a_j} \omega^{(3)} =0, \quad  j=1,\dots, \N.
\lb{a36}
\end{equation}
A normal dtk $\omega_{P_1, P_2}^{(3)}$ associated
with two points $P_1$,
$P_2 \in \hatt \calK_\N$, $P_1 \neq P_2$ by definition
has simple poles at
$P_j$ with residues $(-1)^{j+1}$, $j=1,2$ and
vanishing $a$-periods.  If $\omega_{P,Q}^{(3)}$ is a
normal dtk associated
with $P$, $Q\in\hatt \calK_\N$, holomorphic on
$\calK_\N \setminus \{ P,Q\}$, then
\begin{equation}
\frac{1}{2\pi i} \int_{b_j} \omega_{P,Q}^{(3)} = \int_{Q}^P \omega_j,
\quad  j=1,\dots,\N,
\lb{a37}
\end{equation}
where the path from $Q$ to $P$ lies in
$\hatt \calK_\N$ (i.e.,
does not touch any of the cycles $a_j$, $b_j$). Explicitly, 
one obtains
\begin{align}
\omega^{(3)}_{P_{\infty_+},P_{\infty_-}}&=
-\f{\tilde \pi^\N d\tilde\pi}{y}+\sum_{j=1}^\N d_j\omega_j 
=-\f{\prod_{j=1}^\N (\tilde \pi -\lambda_j)\, d\tilde \pi}{y}, 
\lb{a37a} \\
\omega^{(3)}_{P_1,P_{\infty_+}}&=\f{1}{2}\f{y+y_1}
{\tilde \pi-z_1}\f{d\tilde\pi}{y}-
\f{\prod_{j=1}^\N (\tilde \pi -\hat\lambda_j)\,d\tilde \pi}
{2y}, \lb{a37b} \\
\omega^{(3)}_{P_1,P_{\infty_-}}&=\f{1}{2}\f{y+y_1}
{\tilde \pi-z_1}\f{d\tilde\pi}{y}+
\f{\prod_{j=1}^\N (\tilde \pi -\tilde\lambda_j)\,d\tilde \pi}
{2y}, \lb{a37c} \\
\omega^{(3)}_{P_1,P_2}&=\f{1}{2}\bigg(\f{y+y_1}
{\tilde \pi-z_1}-\f{y+y_2}{\tilde \pi-z_2}\bigg)\f{d\tilde\pi}{y}, 
\quad P_1,P_2\in\calK_\N\setminus\{P_{\infty_+},P_{\infty_-}\}, 
\lb{a37d} 
\end{align}
where $d_j, \lambda_j, \hat\lambda_j,\tilde\lambda_j$, 
$j=1,\dots,\N$ are uniquely determined by the requirement of 
vanishing $a$-periods and we abbreviated $P_j=(z_j,y_j)$, 
$j=1,2$. (If $\N=0$, we use the standard convention that the 
product over an empty index set is replaced by $1$.)

We shall always assume (without loss of generality)
that all poles of
dsk's and dtk's on $\calK_\N$ lie on $\hatt \calK_\N$ (i.e.,
not on $\partial \hatt \calK_n$).

Define the matrix $\tau=(\tau_{j,\ell})_{j,\ell=1,\dots,\N}$ by
\begin{equation}
\tau_{j,\ell}=\int_{b_j}\omega_\ell, \quad j,\ell=1,
\dots,\N. \label{b8}
\end{equation}
Then
\begin{equation}
\Im(\tau)>0, \quad \text{and} \quad \tau_{j,\ell}=\tau_{\ell,j},
\quad j,\ell =1,\dots,\N.  \lb{a18a}
\end{equation}
Associated
with $\tau$ one introduces the period lattice
\begin{equation}
L_\N = \{ \ul z \in\bbC^\N \mid \ul z = \ul m +\tau \ul n,
\; \ul m, \ul n \in\bbZ^\N\}
\lb{a28}
\end{equation}
and the Riemann theta function associated with $\calK_\N$ and
the given homology basis $\{a_j,b_j\}_{j=1,\dots,\N}$,
\begin{equation}
\theta(\ul z)=\sum_{\ul n\in\bbZ^\N}\exp\big(2\pi
i(\ul n,\ul z)+\pi
i(\ul n,\tau \ul n)\big),
\quad \ul z\in\bbC^\N, \label{b9}
\end{equation}
where $(\ul u, \ul v)=\sum_{j=1}^\N \overline{u}_j v_j$
denotes the
scalar product
in $\bbC^\N$. It has the fundamental properties
\begin{align}
& \theta(z_1, \ldots, z_{j-1}, -z_j, z_{j+1},
\ldots, z_\N) =\theta
(\ul z), \lb{a27}\\
& \theta (\ul z +\ul m +\tau \ul n)
=\exp \big(-2 \pi i (\ul n,\ul z) -\pi i (\ul n, \tau
\ul n) \big) \theta (\ul z), \quad \ul m, \ul n \in\bbZ^\N.
\lb{aa51}
\end{align}

Next, fix a base point $Q_0\in\calK_\N\setminus
\Pzeropm,\Pinfpm\}$, denote by
$J(\calK_\N) = \bbC^\N/L_\N$ the Jacobi variety of $\calK_\N$,
and define the
Abel map $\underline{A}_{Q_0}$ by
\begin{equation}
\underline{A}_{Q_0} \colon \calK_\N \to J(\calK_\N), \quad
\underline{A}_{Q_0}(P)=
\big(\int_{Q_0}^P \omega_1,\dots,\int_{Q_0}^P \omega_\N \big)
\pmod{L_\N}, \quad P\in\calK_\N. \label{b10}
\end{equation}
Similarly, we introduce
\begin{equation}
\ul \alpha_{Q_0}  \colon
\Div(\calK_\N) \to J(\calK_\N),\quad
\calD \mapsto \ul \alpha_{Q_0} (\calD)
=\sum_{P \in \calK_\N} \calD (P) \ul A_{Q_0} (P),
\label{aa47}
\end{equation}
where $\Div(\calK_\N)$ denotes the set of
divisors on $\calK_\N$. Here $\calD \colon \calK_\N \to \bbZ$
is called a divisor on $\calK_\N$ if $\calD(P)\neq0$ for only
finitely many $P\in\calK_\N$. (In the main body of this paper 
we will choose $Q_0$ to be one of the branch points, i.e., 
$Q_0\in\calB(\calK_\N)$, and for simplicity we will always choose 
the same path of integration from $Q_0$ to $P$ in all Abelian
integrals.) 

In connection with divisors on $\calK_\N$ we shall employ the
following
(additive) notation,
\begin{multline} \lb{A.17}
\calD_{Q_0\ul Q}=\calD_{Q_0}+\calD_{\ul Q}, \quad \calD_{\ul
Q}=\calD_{Q_1}+\cdots +\calD_{Q_n}, \\
  {\ul Q}=(Q_1, \dots ,Q_n) \in \sigma^n \calK_\N,
\quad Q_0\in\calK_\N,
\end{multline}
where for any $Q\in\calK_\N$,
\begin{equation} \lb{A.18}
\calD_Q \colon  \calK_\N \to\bbN_0, \quad
P \mapsto  \calD_Q (P)=
\begin{cases} 1 & \text{for $P=Q$},\\
0 & \text{for $P\in \calK_\N\setminus \{Q\}$}, \end{cases}
\end{equation}
and $\sigma^n \calK_\N$ denotes the $n$th symmetric product of
$\calK_\N$. In particular, $\sigma^m \calK_\N$ can be 
identified with
the set of nonnegative
divisors $0 \leq \calD \in \Div(\calK_\N)$ of degree $m$.

For $f\in \calM (\calK_\N) \setminus \{0\}$,
$\omega \in \calM^1 (\calK_\N) \setminus \{0\}$ the
divisors of $f$ and $\omega$ are denoted
by $(f)$ and
$(\omega)$, respectively.  Two
divisors $\calD$, $\calE\in \Div(\calK_\N)$ are
called equivalent, denoted by
$\calD \sim \calE$, if and only if $\calD -\calE
=(f)$ for some
$f\in\calM (\calK_\N) \setminus \{0\}$.  The divisor class
$[\calD]$ of $\calD$ is
then given by $[\calD]
=\{\calE \in \Div(\calK_\N)\mid\calE \sim \calD\}$.  We
recall that
\begin{equation}
\deg ((f))=0,\, \deg ((\omega)) =2(\N-1),\,
f\in\calM (\calK_\N) \setminus
\{0\},\,  \omega\in \calM^1 (\calK_\N) \setminus \{0\},
\lb{a38}
\end{equation}
where the degree $\deg (\calD)$ of $\calD$ is given
by $\deg (\calD)
=\sum_{P\in \calK_\N} \calD (P)$.  It is customary to call
$(f)$ (respectively,
$(\omega)$) a principal (respectively, canonical)
divisor.

Introducing the complex linear spaces
\begin{align}
\calL (\calD) & =\{f\in \calM (\calK_\N)\mid f=0
 \text{ or } (f) \geq \calD\}, \;
r(\calD) =\dim_\bbC \calL (\calD),
\lb{a39}\\
\calL^1 (\calD) & =
 \{ \omega\in \calM^1 (\calK_\N)\mid \omega=0
 \text{ or } (\omega) \geq
\calD\},\; i(\calD) =\dim_\bbC \calL^1 (\calD),
\lb{a40}
\end{align}
($i(\calD)$ the index of speciality of $\calD$) one
infers that $\deg
(\calD)$, $r(\calD)$, and $i(\calD)$ only depend on
the divisor class
$[\calD]$ of $\calD$.  Moreover, we recall the
following fundamental
facts.

\begin{theorem} \lb{thm1}
Let $\calD \in \Div(\calK_\N)$,
$\omega \in \calM^1 (\calK_\N) \setminus \{0\}$. Then
\begin{equation}
 i(\calD) =r(\calD-(\omega)), \quad \N\in\bbN_0.
\lb{a41}
\end{equation}
The Riemann-Roch theorem reads
\begin{equation}
r(-\calD) =\deg (\calD) + i (\calD) -\N+1,
\quad \N\in\bbN_0.
\lb{a42}
\end{equation}
By Abel's theorem, $\calD\in \Div(\calK_\N)$,
$\N\in\bbN$ is principal
if and only if
\begin{equation}
\deg (\calD) =0 \text{ and } \ul \alpha_{Q_0} (\calD)
=\ul{0}.
\lb{a43}
\end{equation}
Finally, assume
$\N\in\bbN$. Then $\ul \alpha_{Q_0}
: \Div(\calK_\N) \to J(\calK_\N)$ is surjective
(Jacobi's inversion theorem).
\end{theorem}

Next we introduce
\begin{equation}
\ul W_0=\{0\}\subset J(\calK_\N), \quad 
\ul W_m=\ul \alpha_{Q_0} (\sigma^m \calK_\N), \,\, m\in\bbN 
\lb{a43a}
\end{equation}
and note that while $\sigma^m\calK_\N\not\subset\sigma^n\calK_\N$ 
for $m<n$, one has $\ul W_m\subseteq\ul W_n$ for $m<n$. Thus 
$\ul W_m=J(\calK_\N)$ for $m\geq\N$ by Jacobi's inversion theorem.

Denote by $\ul \Xi_{Q_0}=(\Xi_{Q_{0,1}}, \dots,
\Xi_{Q_{0,\N}})$ the vector of Riemann constants,
\begin{equation}
\Xi_{Q_{0,j}}=\frac12(1+\tau_{j,j})-
\sum_{\substack{\ell=1 \\ \ell\neq j}}^\N\int_{a_\ell}
\omega_\ell(P)\int_{Q_0}^P\omega_j,
\quad j=1,\dots,\N. \lb{aa55}
\end{equation}

\begin{theorem} \lb{thm2}
The set $\ul W_{\N-1}+\ul {\Xi}_{Q_0}\subset J(\calK_\N)$ is the complete 
set of zeros of $\theta$ on $J(\calK_\N)$, that is,
\begin{equation}
\theta (X)=0 \text{ if and only if } X\in\ul W_{\N-1}+\ul {\Xi}_{Q_0} 
\lb{a43b}
\end{equation}
(i.e., if and only if $X=(\ul \alpha_{Q_0} (\calD)+\ul {\Xi}_{Q_0} 
\pmod{L_\N}$ for some $\calD\in\sigma^{\N-1}\calK_\N$). The set 
$\ul W_{\N-1}+\ul{\Xi}_{Q_0}$ has complex dimension $\N-1$.
\end{theorem}

\begin{theorem} \lb{thm3}
Let $\calD_{\ul Q} \in \sigma^n \calK_\N$,
$\ul Q=(Q_1, \ldots, Q_\N)$.  Then
\begin{equation}
1 \leq i (\calD_{\ul Q} ) =s(\leq \N/2)
\lb{a46}
\end{equation}
if and only if there are $s$ pairs of the type
$(P, P^*)\in \{Q_1,
\ldots, Q_\N\}$ (this includes, of course, branch
points for which
$P=P^*$).
\end{theorem}

\begin{remark} \lb{raa19}
While $\theta(\ul z)$ is well-defined (in fact, entire)
for $\ul z\in\bbC^\N$, it is not well-defined on
$J(\calK_\N)=\bbC^\N/L_\N$ because of  \eqref{aa51}.
Nevertheless, $\theta$ is a ``multiplicative
function'' on $J(\calK_\N)$ since the multipliers in
\eqref{aa51} cannot vanish.  In particular, if
$\ul z_1=\ul z_2\pmod{L_\N}$, then $\theta(\ul z_1)=0$ 
if and only
if  $\theta(\ul z_2)=0$.  Hence it is
meaningful to state that $\theta$ vanishes at points of
$J(\calK_\N)$. Since the Abel map
$\ul A_{Q_0}$ maps $\calK_\N$ into  $J(\calK_\N)$, the function
$\theta(\ul A_{Q_0}(P)-\ul{\xi})$ for
$\ul\xi\in\bbC^\N$, becomes a multiplicative function on
$\calK_\N$. Again it makes sense to say that
$\theta(\ul A_{Q_0}(\dott)-\ul{\xi})$ vanishes at points of
$\calK_\N$.
\end{remark}

\begin{theorem} \lb{taa17a} 
Let $\ul Q =(Q_1,\dots,Q_\N)\in \sigma^\N \calK_\N$ and
assume $\calD_{\ul Q}$ to be nonspecial, that is,
$i(\calD_{\ul Q})=0$. Then
\begin{equation}
\theta(\ul {\Xi}_{Q_0} -\ul {A}_{Q_0}(P) + \alpha_{Q_0}
(\calD_{\ul Q}))=0 \text{ if and only if }
P\in\{Q_1,\dots,Q_\N\}. \lb{aa55a}
\end{equation}
\end{theorem}

\begin{lemma} \lb{la6} 
\cite[Lemmas~5.4 and 6.1]{BullaGesztesyHoldenTeschl:1997} Let 
$(x,t),(x_{0},t_{0})\in\Omega$ for some
$\Omega\subseteq\bbR^2$.  Assume
$\psi(\dott,x,t_r)$  to be meromorphic on ${\calK}_n\setminus
\{\Pinfp, \Pinfm, \Pzerop, \Pzerom\}$ with essential singularities
at $\Pinfpm$, $\Pzeropm$ such
that $\tilde \psi(\dott,x,t)$ defined by
\begin{equation}
\tilde \psi (P,x,t) =\psi (P,x,t)
\exp \bigg( i(x-x_0) \int_{Q_0}^P
\Omega_{\infty,0}^{(2)}-i(t-t_{0})\int_{Q_0}^P
\Omega_{0,0}^{(2)}\bigg) \lb{342a}
\end{equation}
is meromorphic on ${\calK}_n$ and its divisor satisfies
\begin{equation}
(\tilde \psi (\dott,x,t))\geq -{\calD}_{\humu (x_{0},t_{0})}.
\lb{342b}
\end{equation}
Here $\Omega_{\infty,0}^{(2)}$ and $\Omega_{0,0}^{(2)}$ are defined
in \eqref{T4.36} and \eqref{T4.37} and the path of integration is
chosen  identical to that in the Abel maps
\eqref{b10} and \eqref{aa47}\footnote{This is to avoid multi-valued 
expressions and hence the use of the multiplicative
Riemann--Roch theorem in the proof of Lemma~\ref{la6}.}. Define a
divisor
${\calD}_0 (x,t)$ by
\begin{equation}
(\tilde \psi (\dott,x,t))={\calD}_0 (x,t)
-{\calD}_{\humu (x_{0},t_{0})}. \lb{342c}
\end{equation}
Then
\begin{equation}
{\calD}_0 (x,t) \in\sigma^n {\calK}_n, \quad {\calD}_0 (x,t) > 0,
\quad \deg ({\calD}_0 (x,t))=n. \lb{342d}
\end{equation}
Moreover, if ${\calD}_0 (x,t)$ is nonspecial for all
$(x,t)\in\Omega$, that is, if
\begin{equation}
i ({\calD}_0 (x,t) ) =0, \quad (x,t)\in\Omega, \lb{342e}
\end{equation}
then $\psi (\dott,x,t)$ is unique up to a constant
multiple (which may depend on $x$ and $t$).
\end{lemma}

\begin{theorem} \lb{taa20}
Suppose $\calD_{\humu}\in\sigma^\N\calK_{\N}$ is nonspecial,
$\humu=(\hmu_{1},\dots,\hmu_{\N})$, and $\hmu_{\N+1}\in\calK_{\N}$
with $\hmu^{*}_{\N+1}\not\in\{\hmu_{1},\dots,\hmu_{\N}\}$.  Let
$\{\hlam_{1},\dots,\hlam_{\N+1}\}\subset\calK_{\N}$ with
$\calD_{\hulam\hlam_{\N+1}}\sim\calD_{\humu\hmu_{\N+1}}$
(i.e., $\calD_{\hulam\hlam_{\N+1}}\in[\calD_{\humu\hmu_{\N+1}}]$).
Then any $\N$ points
$\hnu_{j}\in\{\hlam_{1},\dots,\hlam_{\N+1}\}$, $j=1,\dots,\N$
define a nonspecial divisor $\calD_{\hunu}\in\sigma^\N\calK_{\N}$,
$\hunu=(\hnu_{1},\dots,\hnu_{\N})$.
\end{theorem}
\begin{proof}
Since $i(\calD_{P})=0$ for all $P\in\calK_{1}$, there is nothing to
prove in the special case $\N=1$.  Hence we assume $\N\ge 2$. Let
$P_{0}\in\calB(\calK_{\N})$ be a fixed branch point of $\calK_{\N}$
and suppose that $\calD_{\hunu}$ is special. Then by Theorem
\ref{thm3} there is a pair
$\{\hnu,\hnu^{*}\}\subset\{\hnu_{1},\dots,\hnu_{\N}\}$ such that
\begin{equation}
    \ual_{P_{0}}(\calD_{\hunu})=\ual_{P_{0}}(\calD_{\huunu}),
    \lb{aa100}
\end{equation}
where
$\huunu=(\hnu_{1},\dots,\hnu_{\N})\setminus\{\hnu,\hnu^{*}\}
\subset\sigma^{\N-2}\calK_{\N}$.  Let
$\hnu_{\N+1}\in\{\hlam_{1},\dots,\hlam_{\N+1}\}
\setminus\{\hnu_{1},\dots,\hnu_{\N}\}$ so that
$\{\hnu_{1},\dots,\hnu_{\N+1}\}
=\{\hlam_{1},\dots,\hlam_{\N+1}\}\subset\sigma^{\N+1}\calK_{\N}$. Then
\begin{align}
    \ual_{P_{0}}(\calD_{\huunu\hnu_{\N+1}})
    &=\ual_{P_{0}}(\calD_{\hunu\hnu_{\N+1}})
    =\ual_{P_{0}}(\calD_{\hulam\hlam_{\N+1}}) \no \\
    &=\ual_{P_{0}}(\calD_{\humu\hmu_{\N+1}})
    =-\ua_{P_{0}}(\hmu_{\N+1}^{*})+\ual_{P_{0}}(\calD_{\humu}),
    \lb{aa101}
\end{align}
and hence by Theorem \ref{thm2} and \eqref{aa101},
\begin{equation}
    0=\theta(\uxi_{P_{0}}+\ual_{P_{0}}(\calD_{\huunu\hnu_{\N+1}}))
   =\theta(\uxi_{P_{0}}-\ua_{P_{0}}(\hmu^{*}_{\N+1})
   +\ual_{P_{0}}(\calD_{\humu})). \lb{aa102}
\end{equation}
Since by hypothesis $\calD_{\humu}$ is nonspecial and
$\hmu_{\N+1}^{*}\not\in\{\hmu_{1},\dots,\hmu_{\N}\}$, \eqref{aa102}
contradicts Theorem \ref{taa17a}. Thus,  $\calD_{\hunu}$ is nonspecial.
\end{proof}

{\bf Acknowledgments.} 
It is a great pleasure to dedicate this paper to Sergio Albeverio
on the occasion of his 60th birthday. F.~G. and H.~H.
gratefully acknowledge many years of close interaction with an
extraordinary mentor and friend, who profoundly affected our early
scientific endeavors. 

In addition, we also dedicate this paper with
admiration to Walter E.~Thirring, to honor his  influence on the
physical sciences in general, and on one of us (F.~G.), in particular.



\begin{thebibliography}{10}

\bibitem{Martinezalonso:1984}
L.~Martinez Alonso.
\newblock Soliton classical dynamics in the sine-{G}ordon equation in terms of
  the massive {T}hirring model.
\newblock {\em Phys. Rev. D (3)}, 30:2595--2601, 1984.

\bibitem{BarashenkovGetmanov:1987}
I.~V. Barashenkov and B.~S. Getmanov.
\newblock Multisoliton solutions in the scheme for unified description of
  integrable relativistic massive fields. Non-degenerate $sl(2,\bbc)$ case.
\newblock {\em Comm. Math. Phys.}, 112:423--446, 1987.

\bibitem{BarashenkovGetmanovKovtun:1993}
I.~V. Barashenkov, B.~S. Getmanov, and V.~E. Kovtun.
\newblock The unified approach to integrable relativistic equations: {S}oliton
  solutions over nonvanishing backgrounds. {I}.
\newblock {\em J. Math. Phys.}, 34:3039--3053, 1993.

\bibitem{BarashenkovGetmanov:1993}
I.~V. Barashenkov and B.~S. Getmanov.
\newblock The unified approach to integrable relativistic equations: {S}oliton
  solutions over nonvanishing backgrounds. {II}.
\newblock {\em J. Math. Phys.}, 34:3054--3072, 1993.

\bibitem{BelokolosBobenkoEnolskiiItsMatveev:1994}
E.~D. Belokolos, A.~I. Bobenko, V.~Z. Enol'skii, A.~R. Its, and V.~B. Matveev.
\newblock {\em Algebro-{G}eometric {A}pproach to {N}onlinear {I}ntegrable
  {E}quations}.
\newblock Springer, Berlin, 1994.

\bibitem{Bikbaev:1985}
R.~F. Bikbaev.
\newblock Finite-gap solutions of the massive {T}hirring model.
\newblock {\em Theoret. and Math. Phys.}, 63:577--584, 1985.

\bibitem{BullaGesztesyHoldenTeschl:1997}
W.~Bulla, F.~Gesztesy, H.~Holden, and G.~Teschl.
\newblock Algebro-geometric quasi-periodic finite-gap solutions of the {T}oda
  and {K}ac-van {M}oerbeke hierarchy.
\newblock {\em Mem. Amer. Math. Soc.}, 135(641):1--79, 1998.

\bibitem{ChowdhuryNaskar:1988}
A.~R. Chowdhury and M.~Naskar.
\newblock Monodromy deformation approach to nonlinear equations --- {A} survey.
\newblock {\em Fortschr. Phys.}, 36:939--953, 1988.

\bibitem{Date:1978}
E.~Date.
\newblock On quasi-periodic solutions of the field equation of the classical
  massive {T}hirring model.
\newblock {\em Progr. Theoret. Phys.}, 59:265--273, 1978.

\bibitem{Date:1979}
E.~Date.
\newblock On a construction of multi-soliton solutions of the
  {P}ohlmeyer--{L}und--{R}egge system and the classical massive {T}hirring
  model.
\newblock {\em Proc. Japan Acad. Ser. A Math. Sci.}, 55:278--281, 1979.

\bibitem{Date:1982}
E.~Date.
\newblock Multi-soliton solutions and quasi-periodic solutions of nonlinear
  equations of sine-{G}ordon type.
\newblock {\em Osaka J. Math.}, 19:125--158, 1982.

\bibitem{DavidHarnadShnider:1984}
D.~David, J.~Harnad, and S.~Shnider.
\newblock Multi-soliton solutions to the {T}hirring model through 
the reduction method.
\newblock {\em Lett. Math. Phys.}, 8:27--37, 1984.

\bibitem{DicksonGesztesyUnterkofler:2000}
R.~Dickson, F.~Gesztesy, and K.~Unterkofler.
\newblock A new approach to the {B}oussinesq hierarchy.
\newblock {\em Math. Nachr.}, 198:51--108, 1999.

\bibitem{DicksonGesztesyUnterkofler1:2000}
R.~Dickson, F.~Gesztesy, and K.~Unterkofler.
\newblock Algebro-geometric solutions of the {B}oussinesq hierarchy.
\newblock {\em Rev. Math. Phys.}, 11:823--879, 1999.

\bibitem{FarkasKra:1992}
H.~M. Farkas and I.~Kra.
\newblock {\em Riemann {S}urfaces}.
\newblock Springer, New {Y}ork, second edition, 1992.

\bibitem{Fay:1973}
J.~D.~Fay.
\newblock {\em Theta Functions on Riemann Surfaces}.
\newblock Lecture Notes in Math., Vol. 352, Springer, Berlin, 
1973.

\bibitem{GesztesyHolden:2000}
F.~Gesztesy and H.~Holden.
\newblock {\em Hierarchies of {S}oliton {E}quations and {T}heir
  {A}lgebro-{G}eometric {S}olutions}, monograph in preparation.

\bibitem{GesztesyHolden2:1997}
F.~Gesztesy and H.~Holden.
\newblock A combined sine-{G}ordon and modified {K}orteweg--de {V}ries
  hierarchy and its algebro-geometric solutions.
\newblock {\em In {D}ifferential {E}quations and {M}athematical
{P}hysics, {P}roceedings of an {I}nternational {C}onference held at
the {U}niversity of {A}labama at  {B}irmingham,
{M}arch 16-20, 1999}, {R}.~Weikard  and {G}.~Weinstein, editors,
AMS/IP Studies in Advanced Mathematics, pages 133--173, Amer. Math.
Soc. and International Press, to appear in 2000. 

\bibitem{GesztesyHolden:1999a}
F.~Gesztesy and H.~Holden.
\newblock The classical {B}oussinesq hierarchy revisited.
\newblock {\em Roy. Norw. Soc. Sci. Lett. Trans.}, 1:1--15, 2000.

\bibitem{GesztesyHolden:1999b}
F.~Gesztesy and H.~Holden.
\newblock Darboux-type transformations and hyperelliptic curves. 
\newblock {\em J. reine angew. Math.}, to appear.

\bibitem{GesztesyHolden:1999c}
F.~Gesztesy and H.~Holden.
\newblock Dubrovin equations and integrable systems on hyperelliptic 
curves.
\newblock {\em Math. Scand.}, to appear.

\bibitem{GesztesyRatnaseelan:1996}
F.~Gesztesy and R.~Ratnaseelan.
\newblock An alternative approach to algebro-geometric solutions of 
the {AKNS} hierarchy.
\newblock {\em Rev. Math. Phys.}, 10:345--391, 1998.

\bibitem{GesztesyRatnaseelanTeschl:1996}
F.~Gesztesy, R.~Ratnaseelan, and G.~Teschl.
\newblock The {K}d{V} hierarchy and associated trace formulas.
\newblock In {\em Recent {D}evelopments in {O}perator {T}heory and {I}ts
  {A}pplications}, I.~{G}ohberg, {P}.~{L}ancaster, and {P}.~{N}.~{S}hivakumar,
  editors, {\em Operator {T}heory: {A}dvances and
  {A}pplications}, volume~87, pages 125--163, Birkh\"auser, Basel, 1996.

\bibitem{Holod:1978}
P.~I.~Holod.
\newblock Pseudopotentials and Backlund transformation for Thirring 
equation.
\newblock preprint, 1978 (Russian).

\bibitem{HolodPrikarpatsky:1978}
I.~P.~Holod and A.~K.~Prikarpatsky.
\newblock Classical solutions of two-dimensional Thirring model 
with periodic initial conditions. 
\newblock preprint, 1978 (Russian).

\bibitem{IlievaThirring:1998}
N.~Ilieva and W.~Thirring.
\newblock The {T}hirring model 40 years later.
\newblock Preprint ESI 587, Erwin Sch\"odinger International Institute,
Vienna,
  1998.

\bibitem{KaupLakoba:1996}
D.~J. Kaup and T.~I Lakoba.
\newblock The squared eigenfunctions of the massive {T}hirring model in
  laboratory coordinates.
\newblock {\em J. Math. Phys.}, 37:308--323, 1996.

\bibitem{KaupNewell:1977}
D.~J. Kaup and A.~C. Newell.
\newblock On the {C}oleman correspondence and the solution of the massive
  {T}hirring model.
\newblock {\em Lett. Nuovo Cimento (2)}, 20:325--331, 1977.

\bibitem{KawataMorishimaInoue:1979}
T.~Kawata, T.~Morishima, and H.~Inoue.
\newblock Inverse scattering method for the two-dimensional massive
{T}hirring
  model.
\newblock {\em J. Phys. Soc. Japan}, 47:1327--1334, 1979.

\bibitem{KuznetsovMikhailov:1977}
E.~A. Kuznetsov and A.~V. Mikhailov.
\newblock On the completely integrability of the two-dimensional classical
  {T}hirring model.
\newblock {\em Theoret. and Math. Phys.}, 30:193--200, 1977.

\bibitem{Lee:1993}
J.-H. Lee.
\newblock $n\times n$ {Z}akharov--{S}habat system of the form
  $(d\psi/dx)(z^2-1/z^2){J}\psi+(z{Q}+{P}+{R}/z)\psi$.
\newblock In A.~S. {F}okas, D.~J. {K}aup, A.~C. {N}ewell, and V.~E. {Z}akharov,
  editors, {\em Nonlinear {P}rocesses in {P}hysics}, pages 118--121, 
Springer, Berlin, 1993. 

\bibitem{Lee:1994}
J.-H. Lee.
\newblock Solvability of the derivative nonlinear {S}chr\"odinger equation and
  the massive {T}hirring model.
\newblock {\em Theoret. and Math. Phys.}, 99:617--621, 1994.

\bibitem{Lewittes:1964}
J.~Lewittes.
\newblock Riemann surfaces and the theta function.
\newblock {\em Acta Math.}, 111:37--61, 1964.

\bibitem{Mikhailov:1976}
A.~V. Mikha{\u i}lov.
\newblock Integrability of the two-dimensional {T}hirring model.
\newblock {\em JETP Lett.}, 23:320--323, 1976.

\bibitem{Mumford:1984}
D.~Mumford.
\newblock {\em Tata {L}ectures on {T}heta {II}}.
\newblock Birkh\" auser, Boston, 1984.

\bibitem{NijhoffCapelQuispelLinden:1983}
F.~W. Nijhoff, H.~W. Capel, G.~R.~W. Quispel, and J.~van~der Linden.
\newblock The derivative nonlinear {S}chr\"odinger equation and the massive
  {T}hirring model.
\newblock {\em Phys. Lett. A}, 93:455--458, 1983.

\bibitem{Prikarpatskii:1981}
A.~K. Prikarpatskii.
\newblock Geometrical structure and {B}\"acklund transformations of nonlinear
  evolution equations possessing a {L}ax representation.
\newblock {\em Theoret. and Math. Phys.}, 46:249--256, 1981.

\bibitem{PrikarpatskiiGolod:1979}
A.~K. Prikarpatskii and P.~I. Golod.
\newblock Periodic problem for the classical two-dimensional {T}hirring model.
\newblock {\em Ukrainian Math. J.}, 31:362--367, 1979.

\bibitem{Talalov:1987}
S.~V. Talalov.
\newblock Hamiltonian structure of ``{T}hirring $\times$ {L}iouville'' 
model. Singular solutions. 
\newblock {\em Theoret. and Math. Phys.}, 71:588--597, 1987.

\bibitem{Thirring:1958}
W.~E. Thirring.
\newblock A soluble relativistic field theory.
\newblock {\em Ann. Physics}, 3:91--112, 1958.

\bibitem{TsuchidaWadati:1996}
T.~Tsuchida and M.~Wadati.
\newblock Lax pairs for four-wave interaction systems.
\newblock {\em J. Phys. Soc. Japan}, 65:3153--3156, 1996.

\bibitem{Vaklev:1996}
Y.~Vaklev.
\newblock Soliton solutions and gauge equivalence for the problem of
  {Z}akharov--{S}habat and its generalizations.
\newblock {\em J. Math. Phys.}, 37:1393--1413, 1996.

\bibitem{Villarroel:1991}
J.~Villarroel.
\newblock The {DBAR} problem and the {T}hirring model.
\newblock {\em Stud. Appl. Math.}, 84:207--220, 1991.

\bibitem{WadatiSogo:1983}
M.~Wadati and K.~Sogo.
\newblock Gauge transformation in soliton theory.
\newblock {\em J. Phys. Soc. Japan}, 52:394--338, 1983.

\bibitem{Wisse:1993}
M.~A. Wisse.
\newblock Darboux coordinates and isospectral {H}amiltonian flows for the
  massive {T}hirring model.
\newblock {\em Lett. Math. Phys.}, 28:287--294, 1993.

\end{thebibliography}
\end{document}